\documentclass[apj, numberedappendix]{emulateapj}

\usepackage{graphicx} 

\usepackage{graphicx} 
\usepackage{amsmath} 
\usepackage{amssymb}
\usepackage{exscale, relsize}
\usepackage{amscd}
\usepackage{natbib}

\newcommand{\eg}{e.g.} 
\newcommand{\ie}{\textit{ie.}} 
\newcommand{\viz}{\textit{viz.}} 
\newcommand{\cf}{\textit{cf.}}

\newcommand{\sersic}{S\'ersic}

\newcommand{\zphot}{z_{\textrm{phot}}} 
\newcommand{\zspec}{z_{\textrm{spec}}}

\newcommand{\petrad}{R_{50}} 
\newcommand{\reffz}{R_\mathrm{e}}

\newcommand{\sol}{$_{\odot}$}

\begin{document}

\title{On the Dearth of Compact, Massive, Red Sequence Galaxies in the
Local Universe}

\begin{abstract}
  Using data from the Sloan Digital Sky Survey (SDSS; data release 7),
  we have conducted a search for local analogs to the extremely
  compact, massive, quiescent galaxies that have been identified at $z
  \gtrsim 2$.  We show that incompleteness is a concern for such
  compact galaxies, particularly for {\em low} redshifts ($z \lesssim
  0.05$), as a result of the SDSS spectroscopic target selection
  algorithm.  We have identified 63 $M_* > 10^{10.7} $ M\sol\
  ($\approx 5 \times 10^{10}$ M\sol) red sequence galaxies at $0.066 <
  \zspec < 0.12$ which are smaller than the median size--mass relation
  by a factor of 2 or more.  Consistent with expectations from the
  virial theorem, the median offset from the mass--velocity dispersion
  relation for these galaxies is 0.12 dex.  We do not, however, find
  any galaxies with sizes and masses comparable to those observed at
  $z \sim 2.3$, implying a decrease in the comoving number density of
  these galaxies (at fixed size and mass) by a factor of $\gtrsim
  5000$.  This result cannot be explained by incompleteness: in the
  $0.066 < z < 0.12$ interval, we estimate that the SDSS spectroscopic
  sample should typically be $\gtrsim 75$\% complete for galaxies with
  the sizes and masses seen at high redshift, although for the very
  smallest galaxies it may be as low as $\sim 20$\%.  In order to
  confirm that the absence of such compact massive galaxies in SDSS is
  not produced by spectroscopic selection effects, we have also looked
  for such galaxies in the basic SDSS photometric catalog, using
  photometric redshifts.  While we do find signs of a bias against
  massive, compact galaxies, this analysis suggests that the SDSS
  spectroscopic sample is missing at most a few objects in the regime
  we consider.  Accepting the high redshift results, it is clear that
  massive galaxies must undergo significant structural evolution over
  $z \lesssim 2$ in order to match the population seen in the local
  universe. Our results suggest that a highly stochastic mechanism
  like major mergers cannot be the primary driver of this strong size
  evolution.
\end{abstract}

\author{Edward N Taylor$^{1, 2}$}
\author{Marijn Franx$^1$}
\author{Karl Glazebrook$^3$}
\author{Jarle Brinchmann$^1$}
\author{Arjen van der Wel$^4$}
\author{Pieter G van Dokkum$^5$}

\affil{$^1$ Sterrewacht Leiden, Leiden University, NL-2300 RA Leiden, Netherlands; ent@strw.leidenuniv.nl, 
  $^2$ School of Physics, the University of Melbourne, Parkville, 3010, Australia, 
  $^3$ Centre for Astrophysics \& Supercomputing, Swinburne University of Technology, Hawthorn, 3122, Australia
  $^4$ Max Planck Institut f\"ur Astronomie, D-69117 Heidelberg, Germany, 
  $^5$ Department of Astronomy, Yale University, New Haven, CT 06520-8101}

\shorttitle{On the Dearth of Compact Galaxies in the Local Universe}
\shortauthors{Taylor, Franx, Glazebrook, Brinchmann, van der Wel \&
van Dokkum}



\section{Introduction}

In the simplest possible terms, the na\"ive expectation from
hierarchical structure formation scenarios is that the most massive
galaxies form last.  This is in contrast to the observation that the
bulk of cosmic star formation occurs in galaxies with progressively
lower stellar masses at later times \citep[\eg][]{Juneau, Zheng,
Damen}; the so--called downsizing of galaxy growth.  These
observations have been accommodated within the $\Lambda$CDM framework
with the introduction of a quenching mechanism \citep[e.g.][]{Menci,
Croton, Cattaneo}, which operates to shut down star formation in the
most massive galaxies; this mechanism is also required to correctly
predict the absolute and relative numbers of red galaxies
\citep{DekelBirnboim, Bell2007, Faber2007}.  With this inclusion,
models thus predict that a significant fraction of the local massive
galaxy population should have finished their star formation relatively
early in the history of the universe, with later mergers working to
build up the most massive galaxies.

There is thus a crucial distinction to be made between a galaxy's mean
stellar age, and the time since that galaxy has assumed its present
form \citep[see, e.g.,][]{DeLucia2006}: the most massive galaxies are
expected to be both the oldest {\em and} the youngest galaxies.  They
are the oldest in the sense that their progenitors are expected to
form first in the highest cosmic overdensities---however, these stars
are only assembled into their $z = 0$ configuration relatively
recently.

This leaves (at least) two open questions relating to the quenching of
star formation and the formation and evolution of massive galaxies:
1.)\ When does star formation stop in massive galaxies, and 2.)\ What
happens to galaxies after they have stopped forming stars?

\vspace{0.2cm}

In connection with the first of these questions, deep spectroscopic
surveys have identified massive galaxies with little or no ongoing
star formation at $1 \lesssim z \lesssim 2$
\citep[e.g.][]{Cimatti2004, Glazebrook, McCarthy, Daddi2004}.  At the
same time, color selection techniques like the ERO \citep[and
references therein]{McCarthyEROs}, DRG \citep{Franx-drg}, or BzK
\citep{Daddi2005} criteria have been used to identify massive, passive
galaxies at high redshifts.  While these techniques are deliberately
biased towards certain kinds of galaxies and certain redshift
intervals, advances in techniques for photometric redshift estimation
and stellar population modeling have allowed the selection of
mass-limited samples, and so the construction of representative
samples of the high redshift massive galaxy population
\citep[\eg][]{VanDokkum2006}.

By obtaining very deep rest-frame optical spectra of a
photometric-redshift selected sample of massive galaxies at $z \gtrsim
2$, \citet{Kriek} made a significant advance on previous spectroscopic
and photometric studies.  Of the 36 $\zspec > 2$, $M_* > 10^{11}$
M\sol\ galaxies in the \citet{Kriek} sample, 16 were shown
unambiguously to have evolved stellar populations and little or no
ongoing star formation.  These galaxies also seem to form a red
sequence in $(B-V)$ color, although at low significance
\citep[$3.3\sigma$;][]{Kriek-redseq}.  In other words, these massive
galaxies appear both to have assembled stellar populations similar to
galaxies of comparable mass in the local universe, and to have had
their star formation effectively quenched.

Using Keck laser guide-star assisted adaptive optics and Hubble Space
Telescope imaging, van Dokkum et al.\ (2008, hereafter vD08) measured
sizes for 9 of the 16 strongly quenched galaxies from the
\citet{Kriek} sample.  They found (rest-frame optical) effective radii
in the range 0.5---2.4 kpc; that is, smaller than typical galaxies of
the same mass in the local universe by factors of 3---10.  These
galaxies have stellar mass densities, measured within the central 1
kpc, which are 2---3 times higher than typical local galaxies of the
same mass \citep{Bezanson}.  \citet{Cimatti2008} and Damjanov et al.\
(2009, hereafter D09) have found similarly compact sizes for massive
galaxy samples drawn from $1 < z < 2$ spectroscopic surveys.  Further,
\citet{VanDokkumLetter} have recently measured a velocity dispersion
of $510^{+165}_{-95}$ km/s for one of the galaxies in the vD08 sample,
based on a 29 hr NIR spectrum; this extremely high value is consistent
with the galaxy's measured mass and size. (See also Cappellari et al.\
2009.)

By providing rest-frame optical size measurements for a
representative, mass-limited sample of galaxies
spectroscopically-confirmed to have little or no ongoing star
formation and $z \gtrsim 2$, these results confirm and consolidate the
work of \citet{Daddi2005}, \citet{Trujillo2006}, \citet{Trujillo2007},
\citet{Zirm2007}, and \citet{Toft2007}, as well as $1 < z < 2$ results
from, e.g., \citet{Longhetti} and \citet{Saracco}, and $z \lesssim 1$
results from \citet{VanDerWel2008}.  (See also Buitrago et al., 2008.)

The significance of these results is that while the massive and
largely quiescent galaxies at $z \gtrsim 2$ have stellar populations
that are consistent with their being more or less `fully formed' early
type galaxies, they must {\em each} undergo significant structural
evolution in order to develop into galaxies like the ones seen in the
local universe.  Taken together, these results thus paint a consistent
picture of strong size evolution among massive, early type and/or red
sequence galaxies\footnote{There is considerable, but not total,
overlap between color--selected samples of red sequence galaxies, and
morphology--selected samples of early type galaxies.  While it is
common to use these terms as if they were more or less
interchangeable, it should be remembered that they are not.}---both as
a population and individually---even after their star formation has
been quenched \citep[see also][]{Franx}.  Whatever the mechanism
for this growth in size \citep[see, \eg ,][]{Fan, Hopkins, Naab,
KhochfarSilk}, the formation of massive, passive galaxies is not
monolithic.

\vspace{0.2cm}

The aim of this paper is to test the proposition that there are no
galaxies in the local universe with sizes and masses comparable to
those found at $z \gtrsim 1.5$ --- this is the crux of the argument
against the monolithic formation of massive galaxies.  This work is
based on the latest data products from the Sloan Digital Sky Survey
\citep[SDSS; ][]{York2000, Strauss2002}. In particular, we will focus
on the possibility that such galaxies have been overlooked in SDSS due
to selection effects associated with the construction of the
spectroscopic target sample.

The structure of this paper is as follows: We describe the basic SDSS
data that we have used in \textsection\ref{ch:data}.  In
\textsection\ref{ch:specsample}, we define our sample of compact
galaxy candidates, and present several checks to confirm that these
galaxies are indeed unusually small for their stellar masses.  Then,
in \textsection\ref{ch:select}, we consider the importance of the SDSS
spectroscopic selection for massive, compact galaxies.  In this
Section, we also compare our $z \sim 0.1$ compact galaxy candidates
with the vD08 and \citet{Damjanov} samples.  In Appendix
\ref{ch:photsample}, we provide a complementary analysis in order to
confirm our conclusion that the apparent differences between the high-
and low-redshift samples cannot be explained by selection effects,
including an estimate for the number of compact galaxies that may be
missing from the SDSS spectroscopic sample.  Finally, in
\textsection\ref{ch:discussion}, we compare our results to a similar
studies by \citet{Trujillo} and \citet{Valentinuzzi}, and briefly
examine the properties of our compact galaxies' stellar populations in
comparison to the general $z \sim 0.1$ red sequence galaxy population.
A summary of our main results is given in
\textsection\ref{ch:summary}.  Throughout this work, we assume the
concordance cosmology; \viz: $\Omega_\mathrm{m} = 0.3$,
$\Omega_\Lambda = 0.7$, and $H_0 = 70$ km/s/Mpc.

\section{Basic Data  and Analysis} \label{ch:data}

The present work is based on Data Release 7 \citep[DR7;
][]{Abazajian2009} of the SDSS, accessed via the Catalog Archive
Server\footnote{http://casjobs.sdss.org/CasJobs/} \citep[CAS;
][]{Thakar2008}.  In this section, we describe the basic SDSS data
that we have used, and our analysis of it.  We will search for compact
galaxy candidates in the SDSS spectroscopic catalog; to this end, we
will only consider \texttt{sciencePrimary} objects (a flag indicating
a `science-grade spectrum, and weeding out multiple observations of
individual objects) with either a `star' or `galaxy' photometric
\texttt{type} (\ie , a genuine astronomical source).  The details of
the SDSS spectroscopic sample selection are given in
\citet{Strauss2002}; we will summarize the most relevant aspects of
this process in \textsection\ref{ch:selectfx}.

\subsection{The Basic SDSS Catalog} \label{ch:catalog}

For the basic SDSS catalog, there are two different methods for
performing photometry. The first, the `Petrosian' magnitude, is
derived from the observed, azimuthally averaged (1D) light
profile. The Petrosian radius is defined as the point where the mean
surface brightness in an annulus drops to a set fraction (\viz\ 0.2)
of the mean surface brightness within a circular aperture of the same
radius. Within SDSS, the Petrosian aperture is defined to be twice the
Petrosian radius; this aperture will contain 99 \% of the total light
for a well resolved exponential disk, but may miss as much as 18 \% of
the light for a de Vaucouleurs $R^{1/4}$ profile \citep{Strauss2002,
Blanton-VAGC}.

The second photometric measure is derived from fits to the observed
(2D) distribution of light in each band, using a sector-fitting
technique, in which concentric annuli are divided into 12 30$^{\circ}$
sectors, as described in Appendix A.1 of \citet{Strauss2002}. These
fits are done assuming either an exponential or a de Vaucouleurs
profile, convolved with a fit to the appropriate PSF. For each
profile, the structural parameters (\viz\ axis ratio, position angle,
and scalelength) are determined from the $r$ band image. The more
likely (in a $\chi^2$ sense) of the two profile fits is used to define
`model' magnitudes for each galaxy. For the $ugiz$ bands, these
parameters are then held fixed, and only overall normalization (\ie\
total flux) is fit for.

The basic catalog also provides two different measures of size,
associated with these two magnitude measurements. The Petrosian
half-light radius, $\petrad$, is defined as the radius enclosing half
the `total' Petrosian flux. The catalog also contains best fit
structural parameters, including the effective radius, from a separate
set of fits to each band independently, again for both an exponential
and a de Vaucouleurs profile. Note that whereas the Petrosian
magnitude and size are derived from the observed, PSF-convolved radial
profile, the model values provide a PSF-corrected measure of the
intrinsic size.

We use model magnitudes to construct $ugriz$ SEDs for each object,
since these measurements are seeing--corrected.  From DR7, the basic
SDSS photometric calibration has been refined so that the photometry
is given in the AB magnitude system without the need for any further
corrections \citep{Padmanabhan2008}. For measuring sizes, we will rely
on the best-fit model effective radius, $\reffz$, as determined from
the $z$ band. We also adopt a minimum measured size of $0\farcs75$,
corresponding to half the median PSF FWHM for the SDSS imaging; we
will plot all galaxies with observed sizes smaller than $0\farcs75$ as
upper limits.  (None of our conclusions depend on the choice of this
limit, which ultimately affects only 5 of our lower-mass compact
galaxy candidates.)

\subsection{Derived Quantities}


We have derived rest-frame photometry for each object, based on its
observed $ugriz$ SED and redshift, using the IDL utility InterRest
\citep{Taylor-ecdfs}, with a redshift grid of $\Delta z = 0.001$. In
order to minimize the k-corrections and their associated errors, we
determine rest-frame photometry through the $ugriz$ filters redshifted
to $z = 0.1$, which we denote with a superscript 0.1. We estimate
that the systematic uncertainties are at the level of $\lesssim 0.02$
mag. The agreement between our interpolated rest-frame photometry and
that derived using the SDSS kcorrect algorithm
\citep{Blanton-kcorrect} is very good: our derived $(u-r)$ colors are
$\sim 0.02$ mag bluer for blue galaxies, and $\sim 0.03$ mag redder
for red galaxies.

\begin{figure}[b]
  \includegraphics[width=8.2cm]{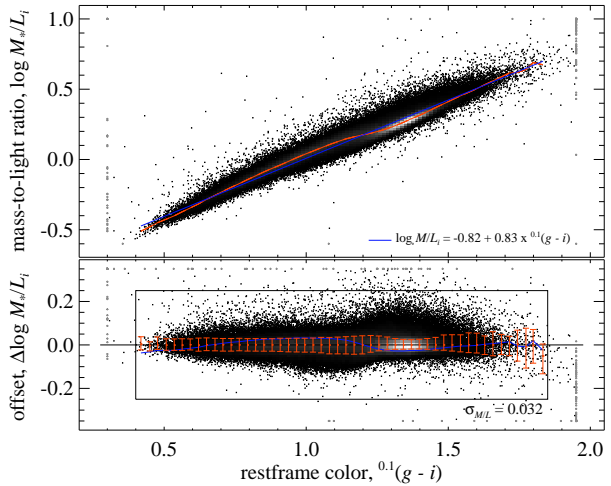}
  \caption{\label{fig:mls}The mass--to--light ratios, $M_*/L_i$, of
     $0.066 < z < 0.12$ galaxies as a function of their $(g-i)$
     color. (Here, $M_*/L_i$ is understood to relate to the $i$ band
     redshifted to $z = 0.1$.)  The greyscale shows the (linear) data
     density in cells, where the data density is high. In the main
     panel, the red line shows the median $M_*/L_i$ in narrow color
     bins; the blue line is a linear fit to these points.  In the
     lower panel, we have simply subtracted away the median relation;
     in this panel, the error bars show the 16/84 percentiles in color
     bins. The simple linear relation shown provides an acceptable
     description of the observed relation, to within 0.02---0.04 dex;
     the global RMS offset from this relation is 0.032 dex. In order
     to avoid selecting `catastrophic failures' in terms of stellar
     mass estimates, we will consider only those galaxies that fall
     within 0.25 dex of the median $M_*/L_i$--$^{0.1}(g-i)$ relation,
     and with $0.4 <\, ^{0.1}(g-i) < 1.8$, as shown by the box in the
     lower panel.}
\end{figure}


\begin{figure*} \centering
  \includegraphics[width=15.5cm]{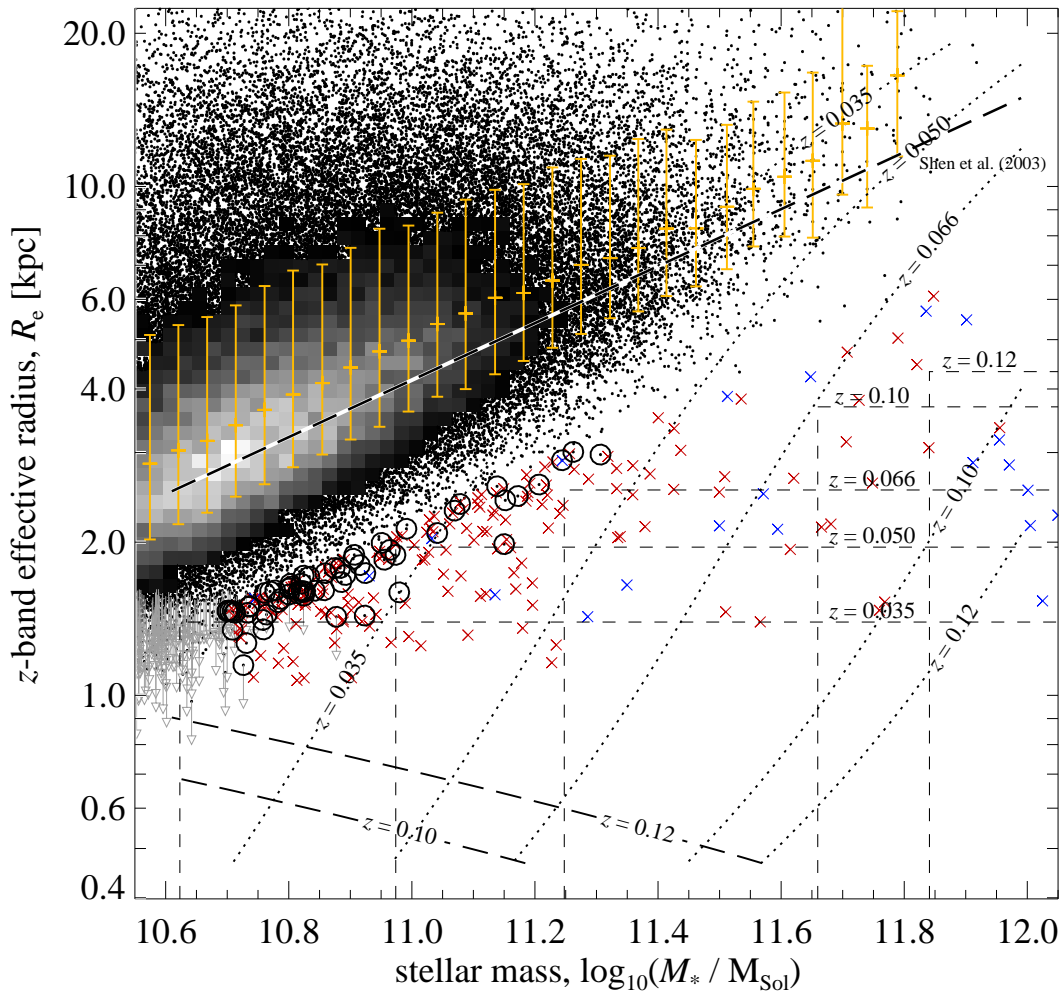}
  \caption{The size--mass relation for massive, red sequence galaxies
    showing the importance of the SDSS spectroscopic selection
    criteria. The points show SDSS galaxies ($0.066 < z < 0.12$)
    selected to have $^{0.1}(u-r) > 2.5$. The yellow points show the
    median size in narrow bins of stellar mass; the error bars show
    the 16/84 percentiles in each bin.  A fit to this median
    size--mass relation for red sequence galaxies is consistent with
    the \cite{Shen} relation for early type galaxies (dashed line),
    albeit offset by 0.05 dex. Individual galaxies that we have
    visually inspected ($M_* > 10^{10.7}$ M\sol ; $\Delta\log \reffz <
    -0.3$ dex) are marked with large symbols.  Galaxies with $M/L$s
    that differ significantly from the main color--$M/L$ relation
    shown in Figure \ref{fig:mls} are marked with small blue crosses.
    Galaxies with obvious problems in their photometry (especially
    those affected by the presence of a bright nearby star or blended
    with other galaxies) are marked with a small red cross; those that
    look okay are plotted as circles.  Galaxies with observed sizes
    smaller than $0\farcs75$ are plotted as upper limits, assuming a
    size of $0\farcs75$. The different lines show how the principal
    selection limits for spectroscopic followup translate onto the
    $(M_*, R_\mathrm{e}$) plane for $z=0.12$, 0.10, 0.066, 0.050, and
    0.35 (top to bottom): the diagonal, long-dashed lines show the
    star/galaxy discriminator; the short-dashed boxes show the
    `saturation' selection limit, and the diagonal dotted lines show
    the`cross-talk' selection limit (see \textsection\ref{ch:selectfx}
    for a detailed discussion). Galaxies lying below these lines will
    not be targeted spectroscopically.
    \label{fig:sizemass}}
\end{figure*}

We make use of stellar mass estimates provided by the MPIA Garching
group.\footnote{Available via http://www.mpa-garching.mpg.de/SDSS/DR7
.} JB has fit the $ugriz$ photometry of all galaxies using the
synthetic stellar population library described by
\citet{Gallazzi2005}, based on \citet{BruzualCharlot2003} models and
assuming a \citet{Chabrier2003} stellar initial mass function (IMF) in
the range 0.1---100 M\sol . The \citet{Gallazzi2005} library contains
a large number of Monte Carlo realizations of star formation
histories, parameterized by a formation time ($1.5 <
t_\mathrm{form}/[\mathrm{Gyr}] < 13.5$), an exponential decay rate ($0
< \gamma/[\mathrm{Gyr}^{-1}] < 1$), and including a number of random
star formation bursts (randomly distributed between $t_\mathrm{form}$
and 0, normalized such that 10 \% of galaxies experience a burst in
the last 2 Gyr). In the fitting, the photometry has been corrected for
emission lines under the assumption that the global emission line
contribution is the same as in the spectroscopic fiber aperture.

The agreement between these SED-fit mass estimates and those of
\citet{Kauffmann2003}, which were derived from spectral line indices,
are excellent: the median offset is -0.01 dex, with a scatter on the
order of 0.1 dex. For the highest masses, however, the SED-fit results
are slightly less robust: for $M_* > 10^{11}$ M\sol , the median
formal error is $\lesssim 0.10$ dex, compared to $\lesssim 0.06$ dex
for the \citet{Kauffmann2003} estimates.

In the upper panel of Figure~\ref{fig:mls}, we show the stellar
mass to light ratios, $M_*/L_i$, for $0.066 < z < 0.12$ galaxies as a
function of their $^{0.1}(g-i) $ color; here again, $L_i$ should be
understood as referring to the $i$-band filter redshifted to $z =
0.1$, or $^{0.1}i$.  Notice that, at least for these mass estimates,
$M_*/L$ is a very tight function of color. In the main panel of this
Figure, the red line shows the median $M_*/L_i$ in narrow color bins.
Making a simple linear fit to these points, we find:
\begin{equation} 
  \log (M_*/L_{i}) = -0.82 + 0.83 \times \, ^{0.1}(g-i) ~ , \label{eq:mls}
\end{equation} 
where both $M_*$ and $L_i$ are in solar units.  (The absolute
magnitude of the sun in the $^{0.1}i$ band is 4.58.)  This relation is
shown in Figure~\ref{fig:mls} as the solid blue line. We present this
relation as an alternative to the popular \citet{BellDeJong} or
\citet{Bell2003} relations.

In the lower panel of Figure~\ref{fig:mls}, we show the dispersion
around the median relation; in this Figure, the error bars show the
16/84 percentiles in bins of color.  Overall, the dispersion around
this relation is just 0.032 dex.  Note that while the simple linear
relation given above provides an acceptable description of the `true'
relation, systematic offsets exist at the 0.02---0.04 dex level.  The
global mean and random offset from this linear relation are 0.002 dex
and 0.040 dex, respectively.

\begin{figure*} \centering
\includegraphics[width=14cm]{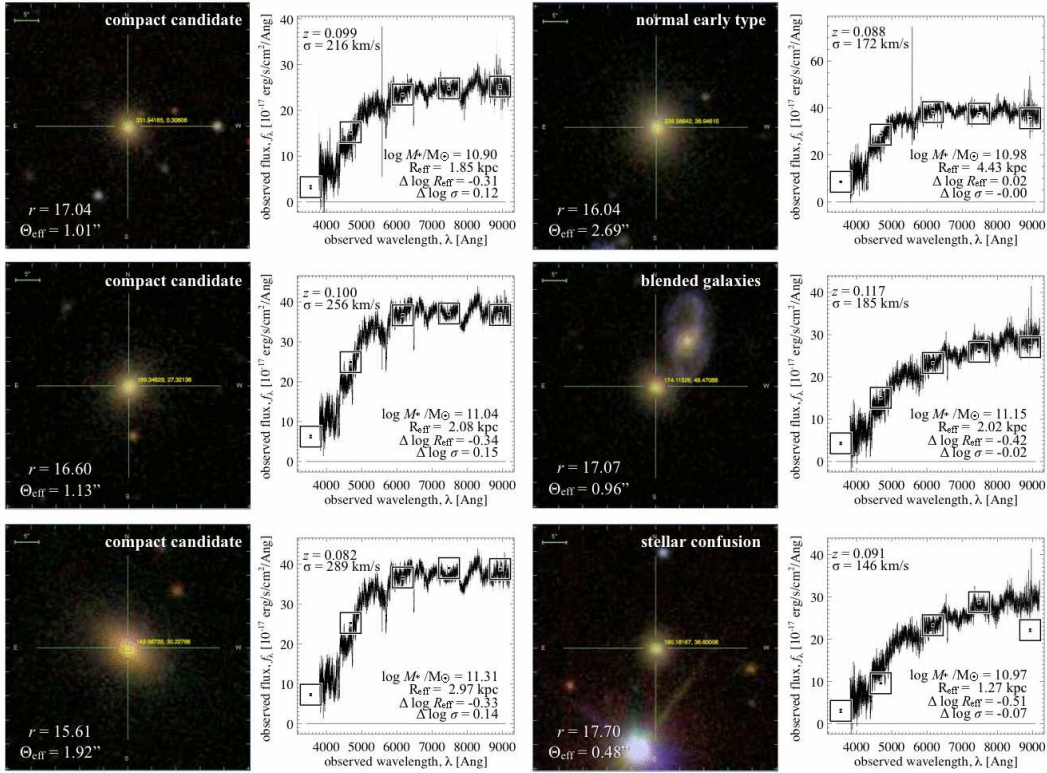}
\caption{\label{fig:imgs}Illustrative examples of the galaxies we
  consider. Clockwise from the top-right, we show a `normal', massive
  early type galaxy that lies very close to the median size--mass and
  velocity dispersion--mass relations, two compact galaxy candidates
  where visual inspection suggested problematic size measurements, and
  three of our compact galaxy candidates.  Each of the three compact
  galaxy candidates shown in this Figure have observed velocity
  dispersions that are approximately consistent with their small
  measured sizes (see \textsection\ref{ch:veldisps}).  For each
  galaxy, we show the SDSS SkyServer thumbnail image used for visual
  inspection, as well as the galaxies' observed spectra; the boxes
  show the SEDs from the photometry, scaled to match the spectroscopic
  flux in the $r$ band.}
\end{figure*}

In both panels, the small grey pluses show points that fall outside
the plotted range.  Notice that there are a small fraction of galaxies
that fall well off the main $M_*/L$--color relation, some by an order
of magnitude or more.  These galaxies also lie significantly off the
main stellar mass--dynamical mass relation and are very likely to
represent catastrophic failures of the stellar mass SED-fitting
algorithm.  This presents a problem when it comes to looking for
outliers in the mass--size plot: selecting the most extreme objects
may well include those objects with the largest errors.  For this
reason, we will restrict our attention to those objects that fall
within 0.25 dex ($\approx 7.8 \sigma$) of the main $M_*/L$--color
relation, and with $0.2 <\, ^{0.1}(g-i) < 1.8$, as shown by the box in
the lower panel of Figure~\ref{fig:mls}.  This selection excludes just
under 600 of the 223292 galaxies shown in Figure~\ref{fig:mls}.

\section{Searching for Massive, Compact, Early-Type Galaxies in the Local Universe} \label{ch:specsample}

\subsection{Identifying Massive, Compact Galaxy Candidates} \label{ch:candidates}

\begin{table*} \begin{center} 
\caption{Properties of our Compact Galaxy Candidates} \label{tab:cat}
\begin{tabular*}{0.975\textwidth}{@{\extracolsep{\fill}}c c c c c c c c c c c c c c c c }
\\ \hline \hline \\
RA & dec & $z$ & $(u-r)_\mathrm{obs}$ & $^{0.1}(u-r)$ & $\log M_*$ & $\Theta_{50, z}$ & $n$ & $\Theta_{50, z}$ & $\reffz$ & $\sigma$ & $T$ & $Z$ &  \\ 
(1)$^a$ & (2)$^a$ & (3)$^a$ & (4)$^a$ & (5) & (6)$^b$ & (7)$^a$ & (8)$^c$ & (9)$^c$ & (10) & (11)$^b$ & (12)$^d$ & (13)$^d$ \\ \\
\hline \hline \\
     190.16652&      13.81563&  0.0855&   2.577&   2.642&  10.700&   0.912&  4.31&   0.917&   1.462&         160&  ... & ... \\
     127.02722&      55.37988&  0.0665&   2.828&   2.999&  10.701&   1.152&  3.26&   1.211&   1.468&         191&    ... & ... \\
     225.31708&      30.58266&  0.0980&   2.858&   2.877&  10.705&   0.808&  3.87&   0.773&   1.464&         195&    ... & ... \\
     227.08531&       7.25325&  0.0764&   2.963&   3.113&  10.709&   0.929&  5.13&   1.188&   1.345&         199&    ... & ... \\
     215.41043&      40.03233&  0.0998&   2.803&   2.813&  10.709&   0.795&  4.11&   0.670&   1.464&         176&   9.775&   0.035&\\
     222.12988&      26.48791&  0.1058&   2.549&   2.540&  10.712&   0.750&  4.65 &  0.722 &   1.453&         155&  ... & ... \\ 
     118.81702&      33.22864&  0.0980&   2.680&   2.697&  10.713&   0.803&  2.67&   0.739&   1.454&         154& -99& -99\\
     143.05707&      11.70454&  0.0811&   2.690&   2.776&  10.726&   0.750&  5.52&   0.830&   1.146&         166&   9.255&   0.132&\\
     204.66577&      59.81854&  0.0704&   2.857&   3.012&  10.731&   0.943&  3.92&   0.974&   1.267&         235&   9.845&   0.229&\\
     230.28553&      24.21978&  0.0809&   2.922&   3.028&  10.733&   0.952&  5.90&   1.298&   1.453&         153&    ... & ... \\
\\ \hline \hline
\end{tabular*} \end{center}
\tablecomments{We give only the first ten candidates here; the
properies of the full sample of 63 galaxies is available as an
electronic table via http://www.strw.leidenuniv.nl/$\sim$ent/. Col.s
(1) and (2): postition in decimal degrees (J2000); Col.\ (3)
spectroscopic redshift; Col.s (4) and (5): observed and rest-frame
colors; Col.\ (6) stellar mass in units of solar masses; Col.\ (7)
apparent De Vaucouleurs effective radius in arcsec; Col.\ (8) and (9)
\sersic\ index and apparent \sersic\ effective radius in arcsec; Col.\
(10) physical De Vaucouleurs effective radius in kpc; Col.\ (11)
velocity dispersion in km/s; Col.\ (12) luminosity-weighted age in
Gyr; Col.\ (13) mean metallicity.  In Col.s (12) and (13), data is only
given for those objects that appear in DR4; values of -99 indicate
unsuccessful fits to the spectra.  \textsc{Sources}. --- $^a$: the
default SDSS \citep{York2000, Strauss2002} catalog for DR7
\citep{Abazajian2009}, accessed via CAS \citep{Thakar2008}; $^b$: the
MPA-JHU catalog for DR7 (accessible via
http://www.mpa-garching.mpg.de/SDSS/DR7/); $^c$: the NYU VAGC
\citep{Blanton-VAGC} for DR7; $^d$: the stellar age and metallicity
catalog of \citet{Gallazzi2005} for DR4.}

\end{table*}

Figure \ref{fig:sizemass} shows the size--mass plot for a sample of
massive, red-sequence galaxies drawn from the SDSS DR7 spectroscopic
sample. These galaxies have been selected to have $^{0.1}(u-r) > 2.5$
and $0.066 < z < 0.12$. These redshift limits have been chosen to
minimize the importance of selection effects and measurement biases,
which we will discuss in \textsection\ref{ch:selectfx}. For now, we
note that, mapping the $r < 17.77$ spectroscopic limit onto $M_*(z)$,
we should be highly complete (volume limited) for $M_* > 10^{10.7}$
M\sol\ and $z < 0.12$.  As a very simple check on this, we note that
for this sample, the median redshift in narrow bins of stellar mass is
within the range $z = 0.098$---0.102 for all $M_* > 10^{10.7}$ M\sol ;
the volumetric center of the $0.066 < z < 0.12$ bin is $z = 0.10$.

The yellow points in this Figure show the median size in narrow bins
of stellar mass; the error bars show the 14/86 percentiles.  For
comparison, the long-dashed line shows the local size--mass relation
for early-type galaxies from \citet{Shen}, corrected for differences
in assumed IMF and cosmology.  Contrary to the findings of
\citet{Valentinuzzi}, a simple fit to the size--mass relation for {\em
red sequence} galaxies ($^{0.1}(u-r) > 2.5$) shown in Figure
\ref{fig:sizemass} is consistent with the \citet{Shen} relation for
{\em early type} ($n > 2.5$) galaxies, albeit offset in size by $0.05$
dex or, equivalently, by $-0.09$ dex in mass.  At fixed mass, the mode
of the distribution is similarly offset (see Figure
\ref{fig:sizedists}); this does not appear to be due to large numbers
of late type galaxies in the sample.

We next select and study very compact galaxies from within the red
sequence sample shown in Figure \ref{fig:sizemass}.  At first glance,
it appears that there may be a few galaxies that lie well below the
main size--mass relation.  However, it must be remembered that by
selecting the most extreme outliers, we will also be selecting those
objects with most egregious measurement errors.

For this reason, we have individually visually inspected all $M_* >
10^{10.7}$ M\sol\ galaxies with inferred sizes that are less than half
the size predicted from the \citet{Shen} relation; \ie\ $\Delta \reffz
< -0.3$ dex.  For sizes smaller than the median relation, the
distribution of sizes around the \citet{Shen} relation is very well
described by a Gaussian with $\sigma = 0.11$ dex; this $\Delta \reffz$
cut thus corresponds to selecting those galaxies whose sizes are
smaller than the mean size (at fixed mass) at the $\gtrsim 2.7 \sigma$
level. (Adopting our own fit to the size--mass relation, this
selection translates to $\Delta \reffz < -0.35$ dex; our results are
otherwise unchanged.) 

We have inspected 280 such objects, and discarded those where there
are obvious reasons to distrust the size measurements.  The most
common reasons for discarding galaxies were confusion with other
galaxies (99 galaxies, including 19 good merger candidates, and two
possible lenses), or with the extended halos, diffraction spikes,
and/or reflections of bright stars (62 galaxies). A further 19
galaxies were clearly disk-like, 5 showed marked asymmetries, and 1
had a very strong AGN spectrum; these candidates were also discarded.
We discarded a further 3 objects with bad or missing data.

In Figure \ref{fig:imgs}, we show several illustrative examples of the
galaxies we are considering.  On the right-hand side of this Figure,
we show a `normal' early type galaxy, with $M_* \approx 10^{11}$
M\sol, which falls very close to the \citet{Shen} relation.  Below
this, we show two of the compact galaxy candidates that we have
rejected on the basis of visual inspection.  On the left-hand side of
this Figure, we show three of the compact galaxy candidates of
different stellar masses that we have retained after visual
inspection.  For each galaxy, we show the thumbnail image from the
SDSS SkyServer\footnote{Also accessible via CAS at
http://cas.sdss.org.}, used for visual inspection.  We also show each
galaxy's observed spectrum and photometry; here, we have scaled the
photometry to match to the integrated $r$-band flux from the observed
spectrum.

In addition to these galaxies with suspect size measurements, we have
excluded a further 27 galaxies whose SED-fit $M_*/L$s are offset from
the main color--$M_*/L$ relation shown in Figure \ref{fig:mls} by more
than 0.25 dex. If we use Equation~\ref{eq:mls} to derive new stellar
mass estimates for these galaxies, all of these galaxies move back
into the main cloud in both Figure \ref{fig:sizemass} and a stellar
mass-dynamical mass plot, with mean/median offsets of $\lesssim 0.02$
dex in both cases.

The 190 galaxies discarded on the basis of inspection are shown in
Figure \ref{fig:sizemass} as small red crosses; the small blue crosses
show the 27 galaxies with discrepant $M_*/L$s.  As a function of
$\Delta \reffz$, the fraction of inspected sources that have been
discarded goes fairly smoothly from 60 \% for $\Delta \reffz \sim -0.3$
dex to $\sim 100$ \% for $\Delta \reffz < -0.5$ dex. The discarded
fraction has a similar dependence on mass: it is $\sim 60$ \% for $M_*
\sim 10^{10.7}$ M\sol , rising to $\sim 85$ \% for $M_* \sim 10^{11}$
M\sol , and 100 \% for $M_* > 10^{11.4}$ M\sol .

This leaves us with a sample of 63 massive, compact, early-type and
red sequence galaxy candidates; these are are marked in Figure
\ref{fig:sizemass} with heavy black circles.  Of those galaxies that
we have retained, 8 \% (5/63) have observed sizes smaller than
$0\farcs75$; all of these have $M_* < 10^{11}$ M\sol .  We have
provided the properties of our compact galaxy candidates in Table
\ref{tab:cat}.

\subsection{Are the Size Measurements Wrong?} \label{ch:sizes}

We have performed a number of checks to validate the small measured
sizes of our compact galaxy candidates. The compact galaxy candidates
do not have significantly larger size measurement errors in comparison
to the full sample shown in Figure \ref{fig:sizemass}. For both the
$r$- and $z$-bands, our candidates are not anomalous in a plot of
Petrosian half-light radius versus model effective radius, nor are
they anomalous in a plot of $r$-band size versus $z$-band size. For
all but two of the candidates, the Petrosian and model magnitudes
agree to within 0.15 mag. The mean offset between model and Petrosian
magnitudes is -0.06 mag for the compact galaxies, compared to -0.08
mag for the full sample shown in Figure \ref{fig:sizemass}. That is,
the compact candidates appear to be well described by the de
Vaucouleurs model fits.

For the New York University (NYU) Value Added Galaxy Catalog (VAGC),
\citet{Blanton-VAGC} have fit the radially-averaged light profiles of
each object, fitting for the \sersic\ index as a free parameter over
the range $0 \le n < 6$.  In order to explore further the issue of the
quality of the de Vaucouleurs profile fits, we have gone to the NYU
VAGC for DR7, and looked up the \sersic\ fit results for each of our
candidates. 

\begin{figure}
	\centering 
	\includegraphics[width=8.6cm]{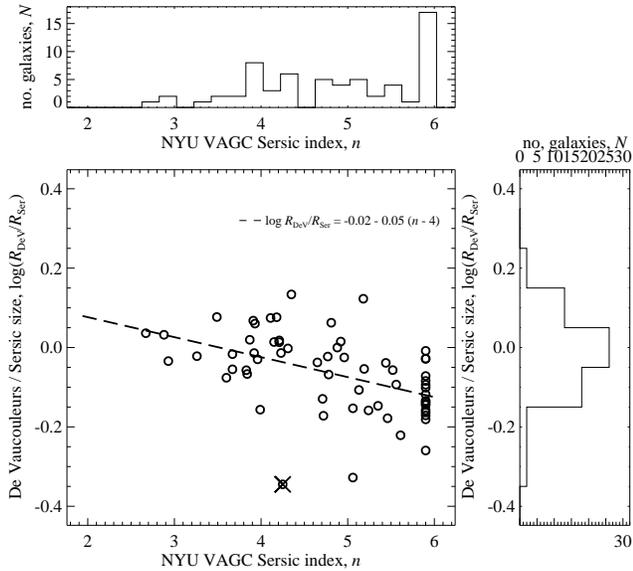}
	\caption{Comparison between effective radii derived assuming a
        de Vaucouleurs ($n = 4$) profile and assuming a \sersic\
        profile ($0 < n < 6$). Whereas the basic SDSS catalog uses a
        sector-fitting technique to fit either an exponential ($n=1$)
        or a de Vaucouleurs ($n=4$) profile, for the NYU VAGC,
        \citet{Blanton-VAGC} have fit the radial profiles of each
        object assuming a general \sersic\ model ($0 < n < 6$). This
        Figure shows the ratio of these two sizes for our compact
        galaxy candidates, based on the $z$-band data, as a function
        of \sersic\ index $n$. Almost all candidates have $n >
        3$---these galaxies are not obviously exponentials. However,
        approximately 25 \% have $n = 5.9$; the maximum value allowed
        in the \citet{Blanton-VAGC} fits.  For these galaxies, the
        median ratio between the two size measurements is 0.88, with
        an RMS scatter of 0.1 dex. In the main panel, we show a
        least-squares fit to the data, the dispersion around this
        relation is $\lesssim 0.1$ dex. \label{fig:sersic}}
\end{figure}

In Figure \ref{fig:sersic}, we show the distribution of \sersic\
parameters for our candidates, as well as a comparison between the
\sersic\ and de Vaucouleurs sizes.  First, we note that nearly all
(59/63) of our compact galaxy candidates have $n > 3$; these are not
late-type (exponential) galaxies. It is therefore unsurprising---but
still reassuring---that the two size measures agree quite well: for
the median galaxy among our candidates, the de Vaucouleurs size is
$\sim 10$\% smaller than the \sersic\ size; the RMS dispersion is 0.10
dex. For comparison, the median quoted error for the de Vaucouleurs
size measurements is 4.6\%.

Notice that about a quarter (17/63) of our candidates have $n = 5.9$
in the NYU VAGC; this is the maximum value allowed in the fits. These
galaxies are considerably more centrally-concentrated than the
canonical de Vaucouleurs $R^{1/4}$-law profile. However, the trend
with increasing \sersic\ index is for the de Vaucouleurs size,
$R_\mathrm{DeV}$, to be systematically lower than the \sersic\ size,
$R_\mathrm{Ser}$: making a least-squares fit to the data shown in
Figure \ref{fig:sersic}, we find $\log R_\mathrm{DeV}/R_\mathrm{Ser} =
-0.02 - 0.05 ~ (n - 4)$.  If these galaxies do have $n > 6$, then we
may well be underestimating their sizes by $\gtrsim 25$ \%.

\citet{Guo} have recently demonstrated that as a result of biases in
the way the background sky level is estimated for the \sersic\
fitting, the NYU-VAGC sizes are systematically underestimated at the
$\gtrsim 15$ \% level for $n \gtrsim 5$.  This problem is
progressively worse for large sizes ($\Theta_\mathrm{e} \gtrsim 1''$)
and bright magnitudes ($r \lesssim 16$); for our compact galaxy
candidates, the effect is likely to be at the $\sim 20$ \% level.  But
note this implies that the difference between the de Vaucoleurs and
\sersic\ sizes is even greater than Figure \ref{fig:sersic} might
suggest: the sizes of the $n \gtrsim 5$ compact galaxies may be
underestimated by as much as $\gtrsim 30$ \%.

As a final check, therefore, we have also re-derived \sersic\
effective radii for our compact galaxy candidates using GALFIT
\citep{Peng} and done a similar comparison as for the NYU VAGC sizes.
The agreement between the GALFIT and VAGC \sersic\ indices is quite
good, with an rms difference in $n$ of 1.1.  Again the vast majority
of objects have $n > 3$.  There are 19 objects that are assigned the
maximum allowed value of $n = 8$, but only 9 of these have $n = 5.9$
in the VAGC.  Making a similar fit to the difference between the
default De Vaucouleurs and the GALFIT \sersic\ effective radii, we
find $\log R_\mathrm{DeV}/R_\mathrm{Ser} = 0.08 - 0.08 ~ (n-4)$.  As
before, we may be underestimating the sizes of high $n$ galaxies by
10---35 \%, although this comparison suggests that we may also be
overestimating the sizes of the few candidates with $n < 4$.  The
median galaxy has a GALFIT \sersic\ effective radius 15 \% smaller
than the default De Vaucouleurs value.  Lastly, we note that there is
a definite mass-dependence to the agreement between the GALFIT
\sersic\ and default De Vaucouleurs effective radii, such that all but
one of the galaxies for which the sizes agree to within 20 \% have
$M_* > 10^{11}$ M\sol .

To summarize the results of this section, then: comparison to 1D and
2D \sersic\ fits does not suggest that the default De Vaucouleurs
effective radii from the SDSS catalog are catastrophically wrong for
our compact galaxy candidates; if anything, these comparisons suggest
that we may in fact be underestimating the sizes of these galaxies by
10---30 \%.

\subsection{A Consistency Check Based on Velocity Dispersions \label{ch:veldisps}}

Assuming that elliptical galaxies are structurally self-similar, the
virial theorem implies that $M_* \propto \reffz \sigma^2$.  At fixed
mass, galaxies with small sizes should therefore have higher velocity
dispersions, with $\Delta \sigma \propto (\Delta \reffz)^{-1/2}$.

\begin{figure}
	\includegraphics[width=8.6cm]{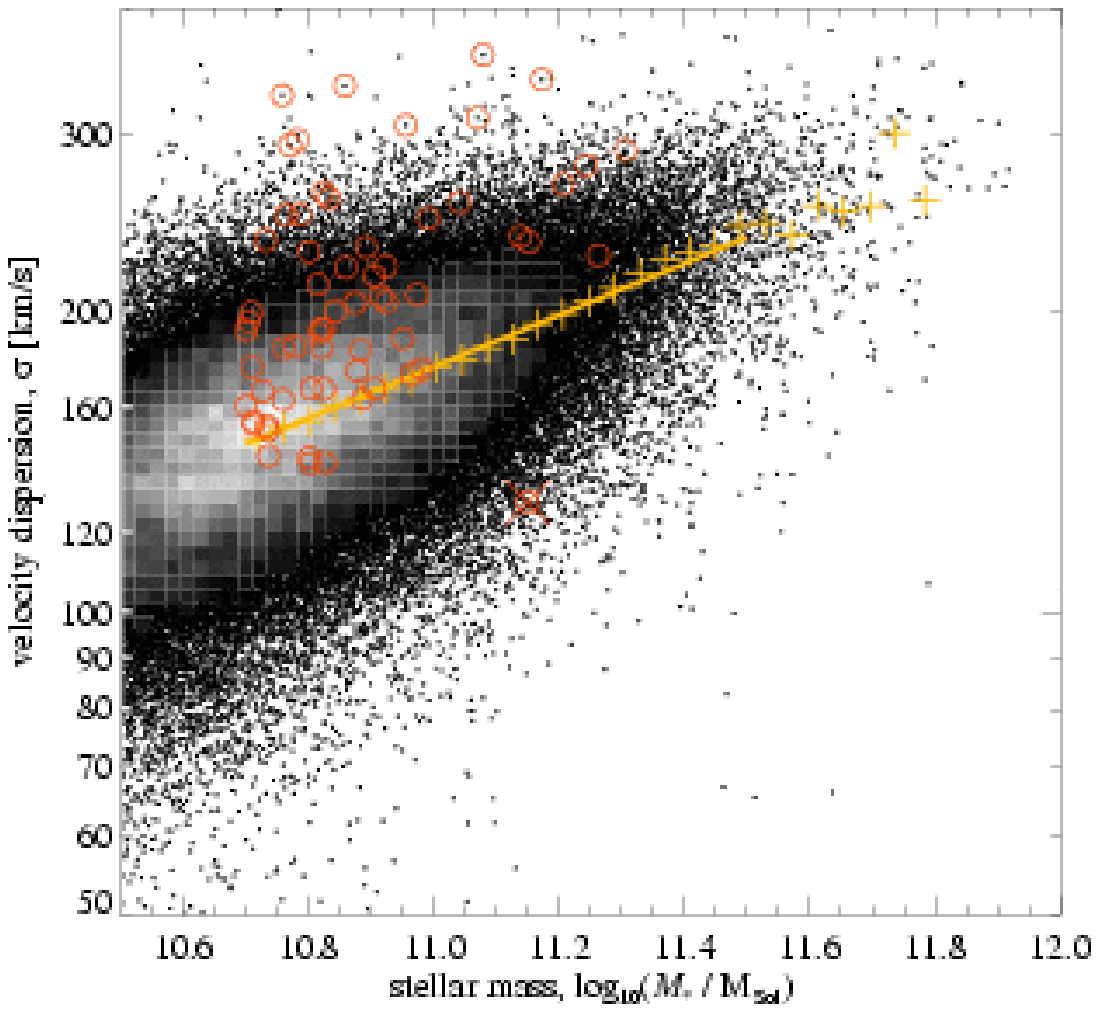}
	\includegraphics[width=8.6cm]{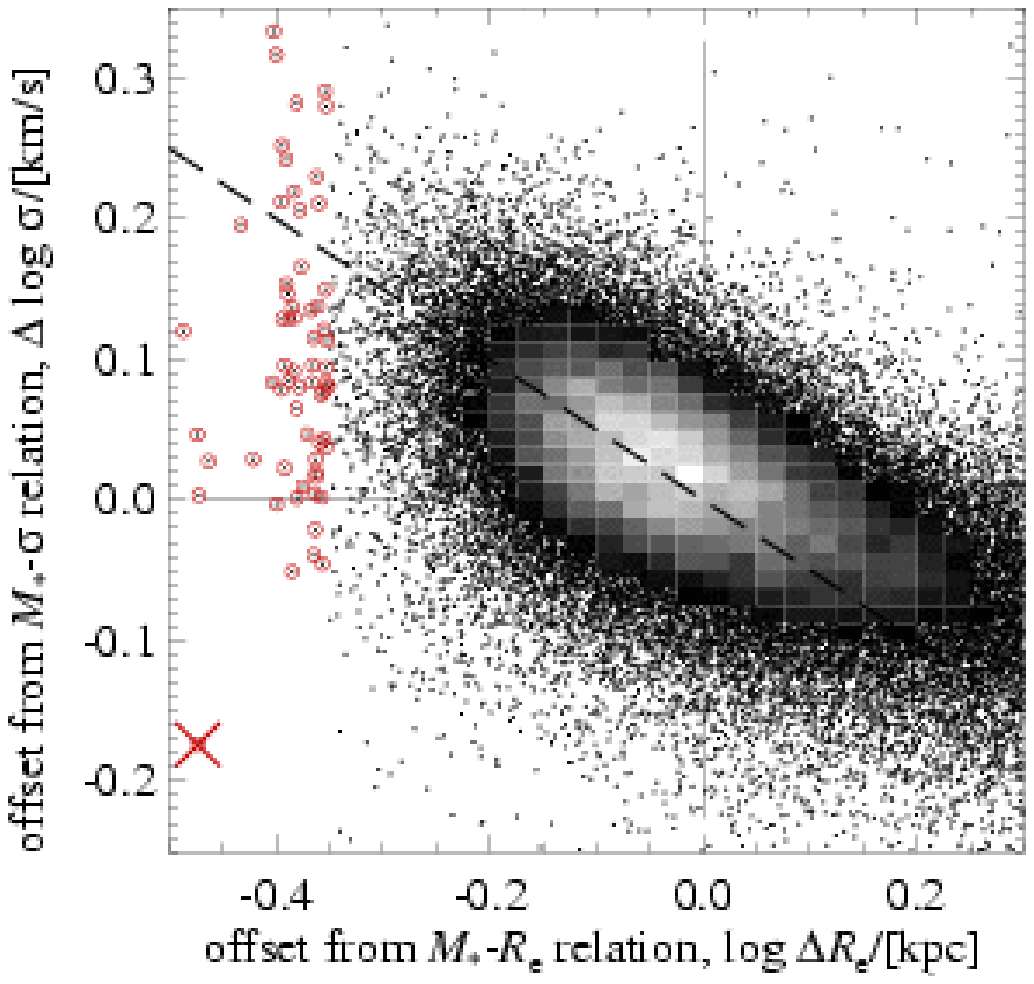}
	\caption{Using observed velocity dispersions to validate the
	measured sizes of our compact galaxy candidates.  {\em Upper
	panel}: the mass--velocity dispersion relation for red
	sequence galaxies at $0.066 < z < 0.12$.  The points and
	greyscale show the SDSS data.  The yellow plusses show the
	median velocity dispersion in narrow bins of stellar mass; the
	solid yellow line shows a simple fit to these points for $10.7
	< \log M_* < 11.5$.  The red circles highlight our compact
	galaxy candidates.  {\em Lower panel}: the offset from the
	$M_*$--$\reffz$ relation, plotted against the offset from the
	$M_*$--$\sigma$ relation for $M_* > 10^{10.7}$ M\sol\ galaxies
	with $^{0.1}(u-r) > 2.5$.  If the offsets from these two
	relations is a function of galaxy size, then we expect
	$\Delta\log\sigma = -0.5 \times \Delta\log\reffz$ (long dashed
	line).  Our compact galaxy candidates are shown as the red
	circles.  In general, the observed velocity dispersions
	support the idea that our compact galaxy candidates are indeed
	relatively small; this is particularly true for those with
	$\sigma > 200$ km/s.  There is one clear exception, marked
	with a cross in both panels; this galaxy is also the most
	extreme outlier in Figure \ref{fig:sersic}.
	\label{fig:deldel}}
\end{figure}

In order to verify that the observed velocity dispersions of our
compact galaxy candidates are consistent with their being genuinely
small, in the lower panel of Figure \ref{fig:deldel} we plot the
offset from the local size--mass relation for early type galaxies,
$\Delta\log\reffz$, against the offset from the $M_*$--$\sigma$
relation, $\Delta\log\sigma$; the $M_*$--$\sigma$ relation itself is
shown in the upper panel of the Figure.  For the lower panel of this
plot, we have shifted the \citet{Shen} relation upwards in size by
0.05 dex to be consistent with the present data set; our conclusions
do not depend on this decision.  The greyscale and points show those
$0.066 < z <0.12$ galaxies with $^{0.1}(u-r) > 2.5$ and $M_* >
10^{10.7}$ M\sol; the red circles indicate those galaxies that we have
identified as compact.

For our compact galaxy candidates, the median offset from the
size--mass relation is $\Delta\log\reffz = -0.38$ dex.  We would
therefore expect a median offset from the $M_*$--$\sigma$ relation
relation of $\Delta\log\sigma = -0.5 \times -0.38 = 0.19$ dex.  The
median value for $\Delta\log\sigma$ is 0.12 dex---roughly 85 \% of the
expected value, and $\sim 1.5$ times greater than the intrinsic
scatter in the relation.  Overall, these results are fairly
consistent, although they do indicate that the sizes may be
underestimated and/or the masses may be overestimated at the level of
10--20 \%.  We note that the difference between the default SDSS and
the NYU VAGC size measurements can account for at least half of this
effect (see \textsection\ref{ch:sizes}).

There is one of our compact galaxy candidates however, whose velocity
dispersion is clearly inconsistent with its being massive and compact,
which we have marked in Figure \ref{fig:deldel} with a cross;
indeed, it has the lowest observed velocity dispersions of all of our
compact galaxy candidates.  This galaxy is also the biggest outlier in
Figure \ref{fig:sersic}.  We will discuss this object in more detail
in \textsection\ref{ch:highz}.

We also note that the observed velocity dispersions of the most
extreme outliers from the size--mass relation ($\Delta\log\reffz
\lesssim -0.4$) are only marginally higher than for galaxies with
`normal' sizes.  Only one of these candidates ($\log M_* = 10.73$) has
$\Delta\log\sigma > 0.18$ dex; the median value of $\Delta\log\sigma$
is 0.03 dex.  It would seem that the effects of `outlier noise' (\ie\
objects being pushed to the edge of the observed distribution by
measurement errors, rather than their true, intrinsic properties)
become dominant at these very extreme values of $\Delta\log\reffz$.

With these caveats, the observed velocity dispersions generally
support the idea that the offsets from both the $M_*$---$\reffz$ and
$M_*$---$\sigma$ relations for our compact galaxy candidates can be
explained by their having small sizes for their masses/velocity
dispersions.

\section{The Importance of Selection Effects for Compact Galaxies}
\label{ch:select}

\subsection{SDSS Spectroscopic Sample Selection} \label{ch:selectfx}

In order to be targeted for SDSS spectroscopic follow-up (and so to
appear in Figure \ref{fig:sizemass}), galaxies have to satisfy a
complicated set of selection criteria \citep{Strauss2002}. In brief,
there is a magnitude cut: objects must be detected at $> 5 \sigma$
significance, and have $r_\mathrm{P} < 17.77$. Any objects that have
been marked as blended and then segmented into smaller objects are
rejected, as are any objects that include saturated pixels, or have
been deblended from objects with saturated pixels. There are also a
series of (low) surface-brightness-dependent criteria that are not
relevant here.

The first important selection criterion for our purposes is the
star/galaxy separation criteria, since we are concerned about bright,
compact galaxies being mistakenly identified as stars. Star/galaxy
separation is done on the basis of the difference between the `PSF'
and `model' magnitudes in the $r$ band. (Here, the PSF magnitude is
derived by fitting the PSF model to each object, in analogy to the
exponential/de Vaucouleurs model fits described in
\textsection\ref{ch:catalog}, and then aperture corrected to $7\farcs4$.)
Specifically, objects are only selected for spectroscopic follow-up
where:
\begin{equation}
  \Delta_\mathrm{SG} \equiv r_\mathrm{PSF} - r_\mathrm{model}
	\ge 0.3 ~ . \label{eq:stargal}
\end{equation}

Further to this star/galaxy discriminator, in order to avoid cross
talk between spectroscopic fibers, galaxies with fiber magnitudes $g <
15$, $r < 15$, and $i < 14.5$ are also rejected.  Lastly, all objects
with $r_\mathrm{P} < 15$ and a Petrosian radius $\Theta_\mathrm{P} <
2''$ are rejected. This criterion was introduced to eliminate ``a
small number of bright stars that that managed to satisfy equation
[\ref{eq:stargal}] during the commissioning phase of the survey, when
the star/galaxy separation threshold was $\Delta_\mathrm{SG} = 0.15$
mag, and was retained for later runs to avoid saturating the
spectroscopic CCDs \citep{Strauss2002}.  \citet{Strauss2002} also note
that of the approximately 240000 $r < 17.77$ objects in runs 752 and
756, none were rejected by the $r_\mathrm{P} < 15$, $\Theta_\mathrm{P}
< 2''$ criterion alone.

In order to model these selections, we need to relate the relevant
observed quantities (\viz, the apparent Petrosian magnitude,
$r_\mathrm{P}$, $gri$ fiber magnitudes, the apparent Petrosian size,
$\Theta_{P}$, and the star/galaxy separation parameter,
$\Delta_\mathrm{SG}$) to intrinsic size and stellar mass.

For a given redshift/distance, the intrinsic size can be trivially
related to the observed effective radius, $\Theta_\mathrm{e}$.  In
order to relate $r_\mathrm{P}$ to $M_*$, we have made a simple fit to
the relation between stellar mass and absolute magnitude in the
observers frame $r$ band (\ie\ with no K-correction) for red sequence
galaxies at $0.066 < z < 0.12$ with $M_r > -21$.  Note that this
method naturally accounts for mass-dependent trends in, \eg ,
metallicity along the red sequence.  The scatter around this relation
is $\sim 0.06$ dex, with no obvious magnitude dependence.  We have
derived similar relations for both $g_\mathrm{P}$ and $i_\mathrm{P}$.

We have also derived empirical relations for $\Theta_\mathrm{P}$,
$\Delta_\mathrm{SG}$, and the difference between the Petrosian and
fiber magnitudes, $\Delta_\mathrm{fib} = m_\mathrm{P} -
m_\mathrm{fiber}$, as functions of $\Theta_\mathrm{e}$ and
$r_\mathrm{P}$, using the sample of massive, red sequence galaxies
shown in Figure \ref{fig:sizemass}. The scatter around these relations
is 0.059 dex (15 \%) , 0.18 mag (18 \%), and 0.11 mag (9 \%)
respectively, with no obvious systematic residuals.

\begin{figure*}
  \centering
  \includegraphics[width=16cm]{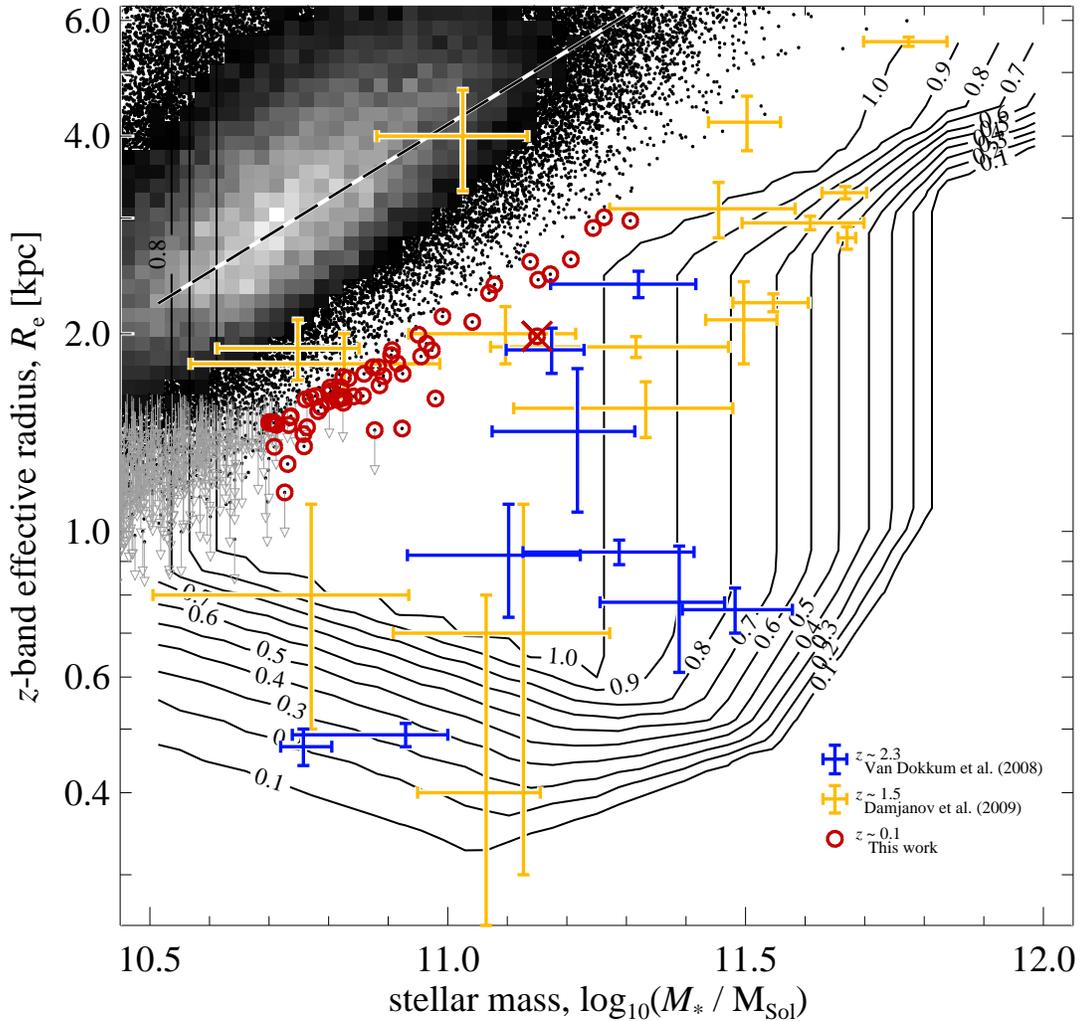}
	\caption{The size--mass relation for massive, red sequence
	galaxies at low and high redshifts. The points and circles
	are for SDSS galaxies, as in Figure \ref{fig:sizemass}; those
	galaxies that we have rejected as per
	\textsection\ref{ch:specsample} are not shown. The contours
	show the relative volume completeness of the SDSS
	spectroscopic sample for $0.066 < z < 0.12$, as marked.  The
	orange points with error bars are the \citet{Damjanov} sample
	of $1.2 < z < 2.0$ galaxies from the GDDS and MUNICS. The blue
	points with error bars are the \citet{VanDokkum} sample of $z
	\sim 2.3$ galaxies from MUSYC. There are no red galaxies in
	the local universe with sizes and masses comparable to the
	compact galaxies found at high redshift. This lack cannot be
	explained by selection effects. \label{fig:sizecomp}}
\end{figure*}

Note that there is a danger of circularity in this argument: any
objects that do not satisfy the selection criteria will not be present
in the sample that we are using to model the selection criteria. The
crucial assumption here, then, is that we can extrapolate the
functions for $\Theta_\mathrm{P}(\Theta_\mathrm{e},~r_\mathrm{P})$,
$\Delta_\mathrm{SG}(\Theta_\mathrm{e},~r_\mathrm{P})$, and
$\Delta_\mathrm{fib}(\Theta_\mathrm{e},~r_\mathrm{P})$ down past the
limits of the spectroscopic sample. In this regard, it is significant
both that the derived functions are smooth all the way down to the
selection limits, and that we do not see obvious cut-offs in the data
associated with these limits.


In Figure \ref{fig:sizemass}, we show how these selection criteria
translate onto the $(M_*, \reffz)$ plane for several example redshifts
between 0.035 and 0.12. The thicker, roughly diagonal, long-dashed
lines show the star/galaxy separation criterion; the dotted lines
show the `cross-talk' fiber magnitude selection; the thinner,
short-dashed boxes show the effect of the `saturation' selection
against bright, compact objects. Note that, for example, a galaxy with
$M_* \gtrsim 10^{11}$ M\sol\ and $\reffz \lesssim 2$ kpc would not be
selected as an SDSS spectroscopic target for $z \lesssim 0.05$.

\subsection{Compact Galaxies at High and Low Redshifts} \label{ch:highz}

In Figure \ref{fig:sizecomp} we again show the size--mass relation for
our sample of massive, red sequence galaxies at $0.066 < z < 0.12$,
with the exception that we have not plotted those galaxies rejected as
per \textsection\ref{ch:candidates}.  Furthermore, in contrast to
Figure \ref{fig:sizemass}, we have used the selection limits derived
in \textsection\ref{ch:selectfx} to estimate the relative completeness
of the SDSS spectroscopic sample across the $0.066 < z < 0.12$ volume;
these are shown by the contours.  These completeness estimates also
include the $r_\mathrm{P} < 17.77$ selection limit, which can be seen
to affect galaxies with $M_* \lesssim 10^{10.6}$ M\sol\ at the distant
end of our redshift window.

For comparison, we have also overplotted the high-redshift samples of
\citet{Damjanov} (yellow points) and \citet[blue]{VanDokkum} (blue
points).  Where we have used size measurements from the $z$-band for
the SDSS galaxies, these high-redshift studies use the NICMOS F160W
filter, which corresponds to rest-frame $r$ at $z=1.6$, moving close
to $g$ by $z=2.3$.  Locally, the difference between $z$- and $r$-band
measured sizes leads to a slightly different slope to the size--mass
relation for red sequence galaxies (a slope of 0.65, rather than
0.56).  The $r$- and $z$- band size--mass relations intersect at
around $M_* \sim 10^{10}$ M\sol ; the mean $r$-band size at $M_* \sim
10^{11}$ M\sol\ is 15 \% larger than in the $z-$band.  That is, by
using $z$-band derived effective radii, we are, if anything, {\em
underestimating} the sizes of the local galaxies in comparison to
those at high redshift.  Similarly, our decision to use the De
Vaucouleurs effective radii given in the basic SDSS catalog, rather
than more general \sersic\ ones appears to lead to an underestimate of
galaxy sizes.  In other words, adopting $r$- or $g$-band derived
sizes, or using \sersic\ instead of De Vaucouleurs effective radii,
would increase the discrepancy between the high- and low-redshift
samples.

\vspace{0.2cm}

There is one of our candidates (marked with a cross) that appears to
have similar properties to one of the larger of the \citet{VanDokkum}
galaxies.  This turns out to be the galaxy whose observed velocity
dispersion is inconsistent with its being genuinely compact
(\textsection\ref{ch:veldisps}); where we would predict
$\Delta\log\sigma = 0.24$ dex, or $\sigma = 310 \pm 70$ km/s, what we
observe is $\Delta\log\sigma = -0.17$ dex and $\sigma = 129 \pm 14$
km/s. This is also the galaxy with the largest difference between the
\sersic-- and De Vaucouleurs--sizes ($\log
R_\mathrm{Dev}/R_\mathrm{Ser} = -0.34$; see
\textsection\ref{ch:sizes}).  Adopting the NYU VAGC \sersic\ size
measurement is not sufficient to reconcile the observed size and mass
with the velocity dispersion: the observed velocity dispersion would
still be too small by $\sim 0.2$ dex.  This galaxy also sits nearly
0.25 dex above the median color--mass-to-light relation shown in
Figure \ref{fig:mls}; using the \citet{BellDeJong} prescription for
$M_*/L$ as a function of $(B-V)$ leads to a stellar mass estimate that
is 0.17 dex lower.  Adopting both this mass estimate and the NYU VAGC
size estimate, we do find consistency between $\Delta\log\reffz$ and
$\Delta\log\sigma$.  In this sense, this galaxy is the weakest of our
compact galaxy candidates---it seems to have had its size
underestimated and/or its mass overestimated.

We also stress that the observed velocity dispersions of the
candidates that lie furthest from the main size--mass relation suggest
that these galaxies have had their sizes significantly underestimated (see
\textsection\ref{ch:veldisps}).

\vspace{0.2cm}

If the \citet{VanDokkum} galaxies were placed at $0.066 < z < 0.12$,
the SDSS spectroscopic completeness would typically be $\gtrsim 75$
\%.  Note, however, that there are two $\reffz < 0.5$ kpc galaxies
from the \citet{VanDokkum} sample and one from the \citet{Damjanov}
sample for which the SDSS completeness is just 20--40 \%.  The average
SDSS completeness for the \citet{VanDokkum} galaxies placed at $0.066
< z < 0.12$ would be 80 \%.

If the \citet{Kriek}/\citet{VanDokkum} galaxies were not to evolve in
either size or number density from $z \sim 2$ to the present day, we
would expect there to be $\sim 6500$ $M_* > 10^{11}$ M\sol\ galaxies
with $\Delta\log\reffz < -0.4$ dex at $0.066 < z < 0.12$, of which
$\sim 5250$ should appear in the SDSS spectroscopic sample.  Instead,
we have only one weak candidate.  


\vspace{0.2cm}

As an interim conclusion, then, we have shown that there are no
galaxies in the local universe (at least as probed by the SDSS
spectroscopic sample) that are directly analogous to the compact
galaxies found at high redshift.  This dearth of compact galaxies
cannot be explained by selection effects.  In Appendix
\ref{ch:photsample}, we confirm this conclusion by searching for
compact galaxy candidates from within the SDSS photometric sample,
using photometric redshifts.  

Moreover, we stress that those galaxies which we have identified as
`compact' are not qualitatively similar to the compact galaxies found
at higher redshifts, which are offset from the local size--mass
relation by at least twice as much again as our local compact galaxy
candidates.

\subsection{The Number Density of Massive, Compact Galaxies} \label{ch:nodens}

\begin{figure} \centering
\includegraphics[width=8.6cm]{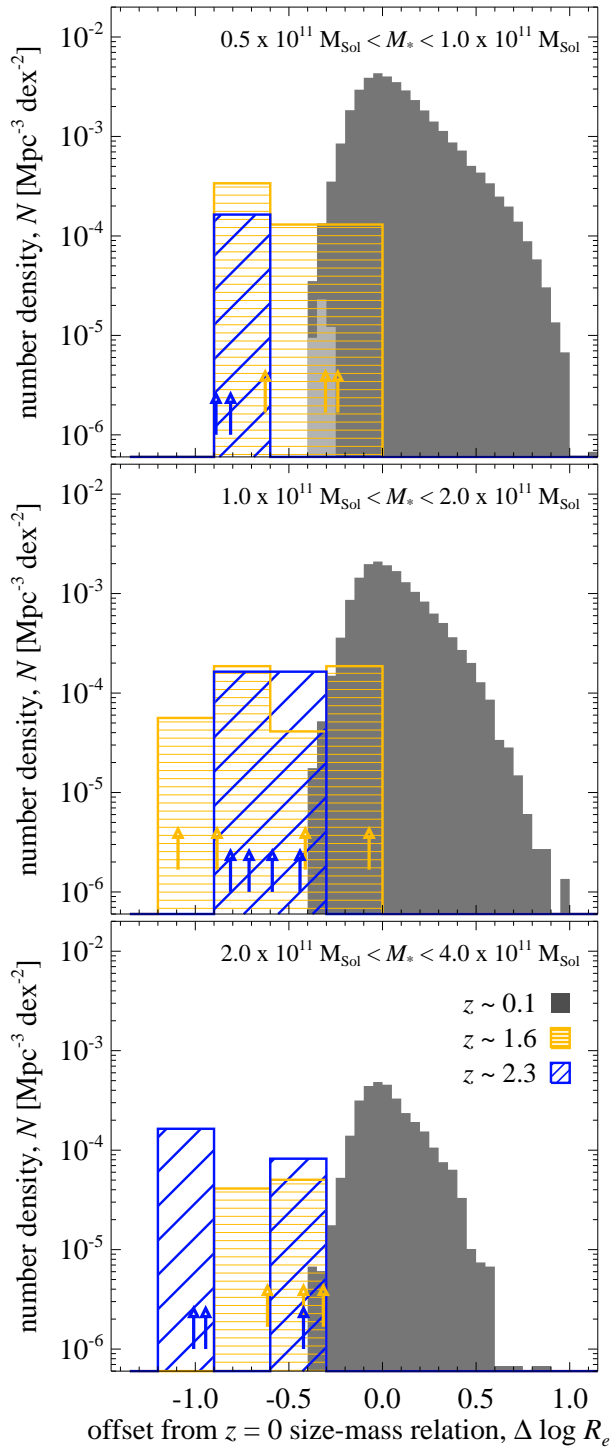}
	\caption{The observed size distribution of massive, red
        galaxies at $z \sim 0.1$, $z \sim 1.6$, and $z \sim 2.3$; each
        panel is for a different mass range as marked. In each panel,
        the solid histogram represents the SDSS spectroscopic sample.
        The blue, diagonally-hatched histogram represents the
        \citet{VanDokkum} sample of nine massive, passive galaxies at
        $z \sim 2.3$; the yellow, horizontally-hatched histogram
        represents the 10 $z \sim 1.6$ GDDS galaxies from
        \cite{Damjanov}.  The arrows at the bottom of each panel
        indicate the positions of the individual high redshift
        galaxies. The $z \sim 2.3$ galaxies have to undergo
        significant structural evolution over $z \lesssim 2.3$ to
        match the properties of local universe galaxies; at least part
        of this evolution has already occurred by $z \sim
        1.6$. \label{fig:sizedists}}
\end{figure} 

In Figure \ref{fig:sizedists}, we provide a more quantitative
statement of our conclusion with respect to the size evolution of
massive galaxies from $z \sim 2$ to $z \sim 0.1$ by plotting the size
distribution for massive, red galaxies in different mass bins.  In
this figure, the filled histograms represent the main SDSS
spectroscopic sample described above.  The horizontal-hatched
histograms show, for comparison, the situation at $z \sim 1.6$, based
on the ten \citet{Damjanov} galaxies drawn from the GDDS; similarly,
the diagonal-hatched histograms show the nine $z \sim 2.3$
\citet{Kriek} galaxies with sizes from \citet{VanDokkum}.

The \citet{Kriek}/\citet{VanDokkum} sample is representative, but not
complete.  In order to derive the densities plotted in Figure
\ref{fig:sizedists}, we have scaled each of the \citet{VanDokkum}
galaxies as follows: first, we have normalized the distribution to
have a density of $1.5 \times 10^{-4}$ Mpc$^{-3}$, which corresponds
to the total number density of all $2 < z < 3$ galaxies to the mass
limit of \citet{Kriek}, derived using the mass function fit given by
\citet{Marchesini}; then, we have scaled this distribution by a factor
of 16/36 to count only those galaxies with little or no ongoing star
formation from \citet{Kriek} that seem to form a red sequence
\citep{Kriek-redseq}.  For the \citet{Damjanov} sample, we are able to
use $1/V_\mathrm{max}$ scalings from \citet{Glazebrook}.

The location of each individual high-redshift galaxy is marked in
Figure \ref{fig:sizedists} with an arrow: the slightly lower blue
arrows show the \citet{VanDokkum} galaxies; the slightly higher yellow
arrows are for the \citet{Damjanov} galaxies.  Clearly, given the
small numbers, the uncertainties on these high redshift values are
quite large, but they do provide a useful order of magnitude estimate
for comparison to the local values.

\vspace{0.2cm}

The clear implication from the comparison between the $z \sim 0.1$ and
$z \sim 2.3$ data in Figure \ref{fig:sizedists} is that, consistent
with the conclusions of \citet{VanDokkum}, not one of the
\citet{VanDokkum} galaxies is consistent with the properties of the $z
\sim 0$ galaxy population.  With the results we have now presented, we
can extend this conclusion by confirming that this discrepancy cannot
be explained by selection effects in the low redshift sample.  

There are local analogs for less than half of the $z \sim 1.6$
galaxies, albeit with considerably higher number densities.  This
would imply that at least some ($\lesssim 50$ \%) of the $z \lesssim
2.3$ evolution has already occurred by $z \sim 1.6$.

\section{Discussion} \label{ch:discussion}

\subsection{Compact Galaxy Properties} \label{ch:props}

\begin{figure*}
	\includegraphics[width=18cm]{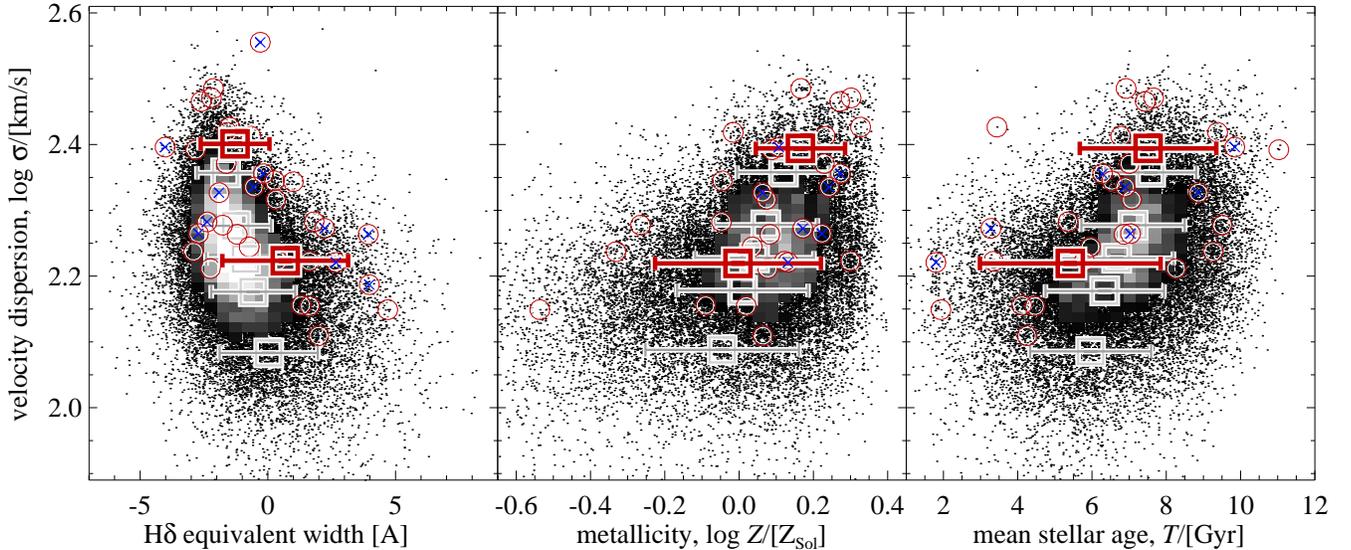}
	\caption{The properties of our compact galaxy candidates in
	comparison to the general population of massive, red sequence
	galaxies.  As in other Figures, the black points and greyscale
	show the data density of all galaxies with $M_* > 10^{10.7}$
	M\sol, $^{0.1}(u-r) > 2.5$, and $0.066 < z < 0.12$; the red
	circles show our compact galaxy candidates.  The grey boxes
	with error bars show the mean and rms scatter in each quantity
	for quintiles of the velocity dispersion distribution; the red
	boxes show the same for those of our compact galaxy candidates
	with $\sigma < 200$ km/s and $\sigma > 200$ km/s separately.
	At fixed velocity dispersion, our compact galaxies have
	slightly lower than average mean ages and slightly higher
	metallicities---however, this result is only significant at the
	$2\sigma$ level.
	\label{fig:props}}
\end{figure*}

In Figure \ref{fig:props}, we plot the properties of our compact
galaxies in comparison to the general massive, red galaxy population.
In each panel, the circles highlight our compact galaxies, while the
points and greyscale show all galaxies with $M_* > 10^{10.7}$ M\sol ,
$^{0.1}(u-r) > 2.5$, and $0.066 < z< 0.12$.  The large grey boxes with
error bars show the mean and standard deviation of each plotted
property in quintiles of the velocity dispersion distribution.
Similarly, the red boxes with error bars show the mean and standard
deviations for our compact galaxy candidates in two bins, separated at
$\sigma = 200$ km/s; the median for this sample.

In each of the panels of Figure \ref{fig:props} (from left to right),
we show the equivalent width of the H$\delta$ line (where negative
values imply emission), the luminosity weighted mean stellar
metallicity, and the luminosity weighted mean stellar age, as derived
from the DR4 SDSS spectra by \citet{Gallazzi2005}.  Because these
estimates are available only for DR4, only around half of our compact
candidates can be plotted in these panels, and only 3/10 of those with
$M_* > 10^{11}$ M\sol .

We have also matched our compact galaxy sample to the AGN sample
described by \citet{Kauffmann-AGN}, for SDSS DR4.  These AGN hosts
have been selected by their [OIII]/H$\beta$ and [NII]/H$\alpha$
emission line ratios; \ie\ the \citet[][BPT]{bpt} diagram.  34 of our
63 galaxies appear in the DR4 catalog; of these, 11 are classified as
AGN on the basis of their emission line ratios.  This is slightly
higher than the AGN fraction of the parent sample, which is in the
range 20---26 \% for the mass range we are considering.  Of the 11
galaxies identified as AGN hosts, five sit on or slightly above the
main $M_*$---$L_\mathrm{[OIII]}$ relation, with $L_\mathrm{[OIII]}
\approx 10^6$ L\sol , four have $L_\mathrm{[OIII]} \sim 10^{7-8}$
L\sol , and one is quite high luminosity, with $L_\mathrm{[OIII]} =
10^{8.7}$ L\sol .  These 11 galaxies are marked in each panel of
Figure \ref{fig:props} with a small blue cross.

\citet{Kauffmann-AGN} also provide revised stellar mass and velocity
dispersion measurements for these galaxies. Accounting for the
presence of an AGN does not have a major impact on these measurements:
the masses and velocity dispersions change at the level of 0.05 dex
and 16 km/s, respectively. That is, while it is possible that an
optically bright point source may bias the measured sizes of these
galaxies downwards, within the stated errors, the AGN does not
significantly affect the derived values of $M_*$ or $\sigma$.  (It is
relevant here that only one of our compact galaxy candidates shows a
significant residual point-source after subtracting off the best-fit
\sersic\ profile, as produced by GALFIT; see
\textsection\ref{ch:sizes}.)

\vspace{0.2cm}

Looking now at Figure \ref{fig:props}, it is clear that the majority
of our compact galaxy candidates have quite old stellar populations.
For the $\sigma < 200$ km/s bin, the median age is 6 Gyr, although the
ages do range from 2 to 10 Gyr, while all but one of the $\sigma >
200$ km/s candidates have $T > 6$ Gyr.  Among the lower velocity
dispersion candidates, there is a clear tendency towards relatively
high equivalent widths for H$\delta$ absorption, suggestive of a
relatively recent ($\lesssim 2$ Gyr) star formation event.

At fixed velocity dispersion, our compact galaxy candidates may have
slightly higher metallicities, and be slightly younger than average.
Using bootstrap-resampling on similar sized samples drawn randomly
from the mass-limited sample, and controlling for velocity dispersion,
these results are weakly significant at best: $1.9 \sigma$ for the
age, and $< 1\sigma$ for metallicity.  Considering only the higher
velocity dispersion candidates ($\sigma > 200$ km/s), the significance
of these differences become $1.9 \sigma$ and $2.2 \sigma$ for the age
and metallicity offsets, respectively.  This weakly significant result
should be contrasted with the results of \citet{Shankar} and
\citet{VanDerWel2009}, who find that, on average and at fixed
dynamical mass, early type galaxies with higher velocity dispersions
(or, equivalently, smaller sizes) have older mean stellar ages.

While the younger mean stellar ages and lower metallicities of our
compact galaxy sample are only weakly significant, both would imply a
relatively late start to star formation for these galaxies and/or
their progenitors.  But if these galaxies grow in size through mergers
(for example) then it is possible that these galaxies are small not
because their formation is delayed relative to other galaxies of the
same mass or velocity dispersion, but rather because they have had
fewer mergers overall, or perhaps just fewer recent mergers.  That is,
it may be that, at fixed mass, these compact galaxies are in fact {\em
older}, in the sense that they have been assembled earlier, and
existed in (more or less) their present form for longer than other
galaxies of the same mass or velocity dispersion.

\subsection{Comparison to Other Recent Works}

In a similar study to this, using sizes and photometry from the NYU
VAGC for SDSS DR6, \citet{Trujillo} have recently reported the
detection of 29 $z < 0.2$ galaxies with $M_* > 8 \times 10^{10}$
M\sol\ and $R_\mathrm{e} < 1.5$ kpc.  In contrast, we find just one
galaxy from our $0.066 < z < 0.12$ red sequence galaxy sample that
satisfy these mass and size criteria; this implies a difference in
volume densities of a factor of 5.5.  Most of this difference is
explained by the fact that we have preselected our compact galaxies to
be red.  Of the \citet{Trujillo} galaxies, only around 30 \% (9/29)
satisfy our $^{0.1}(u-r) > 2.5$ criterion, bringing our number
densities into agreement.  On the other hand, if we also look at
$^{0.1}(u-r) < 2.5$ galaxies, inspected as per
\textsection\ref{ch:candidates}, we find only 7 additional candidates,
only 3 of which have $M_* > 8 \times 10^{10}$ M\sol ; the most massive
of these blue compact galaxy candidates is $7.2 \times 10^{10}$ M\sol .

We also note that the \citet{Trujillo} galaxies have considerably
smaller observed sizes than the galaxies we consider here.  (Recall
that we adopt a minimum observed size of $0\farcs75$; for galaxies
with inferred sizes smaller than this, we adopt a size of $0\farcs75$
as an upper limit on the true size.)  The largest observed size among
the \citet{Trujillo} sample is $0\farcs70$; the median is just
$0\farcs48$.  Enforcing our minimum allowed size of $0\farcs75$, only
one of the \citet{Trujillo} galaxies, irrespective of color, would
have $\reffz < 1.5$ kpc; the median size for the sample would become
2.1 kpc.  It is also relevant here that only one of the
\citet{Trujillo} galaxies is at $z < 0.12$; the other 28 are all found
at $0.12 < z < 0.20$.  This suggests that the \citet{Trujillo} size
measurements may be biased by inadequate resolution.

The key difference between our compact galaxy sample and that of
\citet{Trujillo} is that they find a median velocity dispersion which
is only 0.04 dex higher than their control sample, even though the
mean size and mass are offset from the \citet{Shen} relation by $-0.5$
dex.  This discrepancy can only be explained by either very large
structural differences, or if the \citet{Trujillo} sample is
disproportionally affected by large measurement errors in size and/or
mass.  In this context it is significant that, the observed velocity
dispersions for our candidates with $\Delta\log\reffz < -0.4$ dex
imply that their very small inferred sizes are produced by large
errors in the measured sizes (see \textsection\ref{ch:veldisps}).
Only with followup observations will we be able to determine the true
nature of the \citet{Trujillo} galaxies.  In any case,
\citet{Trujillo} also do not find any galaxies directly comparable to
those found at $z \gtrsim 2$.

\vspace{0.2cm}

Even more recently, \citet{Valentinuzzi} have described a sample of
147 compact galaxies selected from the WIde-field Nearby
Galaxy-cluster Survey (WINGS) of X-ray selected clusters at $0.04 < z
< 0.07$.  Unlike in this work, \citet{Valentinuzzi} do find galaxies
with properties comparable to the 3 largest \citet{VanDokkum}
galaxies; similarly, there are local WINGS analogs for 8 of the 10
GDDS galaxies from \citet{Damjanov}.  However, this relies on their
scaling the high redshift galaxies' masses down by 0.15 dex to account
for the poor treatment of NIR-luminous thermally pulsating asymptotic
giant branch (TPAGB) stars in the \citet{BruzualCharlot2003} stellar
population models.  While both \citet{Kriek2009} and \citet{Muzzin}
show that the stellar masses for the $z \sim 2.3$ implied by different
models vary by $\sim 0.1$ dex, we have not applied such a correction
here.

We note, however, that the high- and low-redshift samples have been
treated consistently here, including the fact that all masses were
derived from the rest-frame optical.  Further, we note that
\citet{VanDerWel2006} have shown that stellar masses derived from the
rest-frame optical and using \citet{BruzualCharlot2003} models are
consistent with the dynamical masses of $z < 1$ galaxies, and are
unaffected by the TPAGB uncertainties.

The \citet{Valentinuzzi} compact galaxies sample is selected by
effective surface mass density, $\Sigma_\mathrm{e} = M_* / 2 \pi
\reffz^2 > 4 \times 10^{9}$ M\sol\ kpc$^{-2}$, in the range 3---50
$\times 10^{10}$ M\sol.  Our compact galaxy selection is roughly
equivalent to $\Sigma_\mathrm{e} \gtrsim 3.6 \times 10^{9}$ M\sol\
kpc$^{-2}$; nearly half (28/63) of our compact galaxy candidates
satisfy the \citet{Valentinuzzi} $\Sigma_\mathrm{e}$ criterion.  For
their sample, \citet{Valentinuzzi} derive a number density of $1.6
\times 10^{-5}$ Mpc$^{-3}$; to our mass limit of $M_* > 10^{10.7}$
M\sol , this value becomes $1.2 \times 10^{-5}$ Mpc$^{-3}$.  These
values are solid lower limits, as they assume that no such galaxies
exist outside of the clusters observed by WINGS.  For our sample,
however, the number density of $M_* > 10^{10.7}$ M\sol\ galaxies with
$\Sigma_\mathrm{e} > 4 \times 10^{9}$ M\sol\ kpc$^{-3}$ is just $2.85
\times 10^{-7}$ Mpc$^{-3}$.

That is, after correcting as best we can for the different stellar
mass limits, our number densities are inconsistent by a factor of more
than 40 with those found by \citet{Valentinuzzi}.  Again, our use of
$z$-band effective radii leads to smaller measured sizes than for
bluer bands; this discrepancy would only increase using $r$- or $g$-band
measured sizes.  Either our results are badly affected by unexplained
selection effects, or there are large discrepancies between our size
and mass estimates and those of \citet{Valentinuzzi}.

We have considered possible spectroscopic selection effects that could
bias against bright, compact objects in \textsection\ref{ch:selectfx},
and shown these to be relevant for $z \lesssim 0.05$.  These effects
may well explain why \citet{Valentinuzzi} were able to match only a
small fraction of their compact galaxies (which have $0.04 < z <
0.07$) to objects in the (DR4) SDSS spectroscopic catalog.  We have
shown, however, that our $0.065 < z < 0.12$ results are not strongly
affected by these kinds of selection effects (Figure
\ref{fig:sizecomp}, see also Appendix A).  Our estimated completeness
is more than 60 \% for all galaxies in the \citet{Valentinuzzi} sample
and greater than 90 \% for 90 \% of the sample.  The selection effects
considered in \textsection\ref{ch:selectfx} thus cannot explain the
difference in our inferred number densities.

An alternative explanation is that the \citet{Valentinuzzi} galaxies
only exist in rich clusters, and that SDSS suffers much higher
spectroscopic incompleteness in such dense fields because of fiber
collisions.  A completely indepdendent estimate can be obtained from
the \citet{Faber} sample: we find that 5/319 of these galaxies have
sizes smaller than the mass--size relation by a factor of 2 or more.
This fraction for clusters that is approximately 15 times higher than
what we find for all galaxies in SDSS.  While not conclusive, this
does suggest that SDSS may suffer from additional incompleteness
beyond the effects we consider here.  That said, we note that several
studies \citep[e.g.][]{Kauffmann2004, Park, Weinmann} have found
little or no evidence for an environmental dependence of the
size--mass relation within SDSS.

Thus we can find no easy explanation for the difference between the
\citet{Valentinuzzi} results and our own.  Here again, velocity
dispersion measurements would provide an useful consistency check on
the \citet{Valentinuzzi} size and mass measurements.

Even despite these differences, however, we note that
\citet{Valentinuzzi} conclude that---barring large systematic errors
in the high-redshift measurements---at least 65 \% (\cf\ our value of
100\%) of the $z \sim 2.3$ galaxies from \citet{VanDokkum} and at
least 20 \% (\cf\ our value of 60 \%) of the $z \sim 1.6$ galaxies
from \citet{Damjanov} have disappeared from the local universe.
Accepting the high-redshift results, these galaxies simply cannot
evolve passively and statically into the red sequence and/or early
type galaxies found in the local universe.

\section{Summary and Conclusions} \label{ch:summary}

The central question of this work has been the existence or otherwise
of massive, compact, quiescent and/or early type galaxies in the local
universe, and particularly the importance of selection effects in the
SDSS spectroscopic sample for such galaxies.  We have shown that,
especially for lower redshifts ($z \lesssim 0.05$), galaxies with the
masses and sizes of those found at $z \gtrsim 2$ would not be targeted
for spectroscopic followup (Figure \ref{fig:sizemass}).  The main
reason for this is not the star/galaxy separation criterion, but
rather the exclusion of bright and compact targets in order to avoid
saturation and cross-talk in the spectrograph (see
\textsection\ref{ch:selectfx}).

We have therefore conducted a search for massive, compact galaxies at
$0.066 < z < 0.12$, where these selection effects should be less
important.  We estimate that for $0.066 < z < 0.12$, the average
completeness for galaxies like those from \citet{VanDokkum} and
\citet{Damjanov} would be $\gtrsim 25$ \% at worst, and $\sim 80$ \%
on average (Figure \ref{fig:sizecomp}).

Starting from a sample of massive ($M* > 10^{10.7}$ M\sol) red
sequence ($^{0.1}(u-r) > 2.5$) galaxies, we have selected the 280
galaxies with inferred sizes that are a factor of 2 or more smaller
than would be expected from the \citet{Shen} $M_*$--$\reffz$ relation
for early type galaxies.  In order to confirm their photometry and
size measurements, we have visually inspected all of these objects.
Unsurprisingly, by selecting the most extreme outliers, a large
fraction of these objects ($\sim 70$\%) appear to be instances where
the size and/or stellar mass estimates are unreliable
(\textsection\ref{ch:candidates}).

For the 63 galaxies with no obvious reason to suspect their size or
stellar mass estimates, there is good agreement between the default
SDSS size measurement (based on the 2D light distribution, using a
sector-fitting algorithm, and assuming a de Vaucouleurs profile), and
those given in the NYU VAGC (based on the azimuthally average growth
curve, assuming a more general \sersic\ profile).  However,
particularly for galaxies with high $n$, the de Vaucouleurs size
measurement is systematically smaller than the \sersic\ one, at the
level of $\lesssim 25$ \% (\textsection\ref{ch:sizes}).

In general, and as expected, our 63 compact galaxy candidates have
significantly higher than average velocity dispersions (Figure
\ref{fig:deldel}).  While it remains possible that the sizes of at
least some of our compact galaxy candidates may have had their sizes
underestimated by $\sim 30$ \%, in general, the relatively high
observed velocity dispersions support the notion that they are indeed
unusually compact given their stellar masses
(\textsection\ref{ch:veldisps}).

\vspace{0.2cm}

Among our compact galaxy candidates, there are no galaxies with sizes
comparable to those found $z \sim 2.3$ by \citet{VanDokkum}; we find
analogs for $\lesssim 50$ \% of the \citet{Damjanov} galaxies at $z
\sim 1.6$ (Figures \ref{fig:sizecomp} and \ref{fig:sizedists}).  This
lack cannot be explained by selection effects.  To confirm this, we
have also compared the size--mass diagram, constructed using
photometric redshifts, based on both the full photometric sample and
the spectroscopic sub-sample (Appendix \ref{ch:photsample}).  While it
is conceivable that SDSS is missing a few massive, compact galaxies,
there are again no signs of galaxies comparable to those of
\citet{VanDokkum} or \citet{Damjanov}.

It is not impossible that some systematic errors in the
estimation of $M_*/L$s for the high redshift galaxies (\eg , an
evolving stellar IMF) mean that their stellar masses are vastly
overestimated, however it would require an overestimate of $\gtrsim
0.7$ dex to reconcile the \citet{VanDokkum} galaxies with the
sizes of the smallest galaxies we have identified in the SDSS catalog.

Accepting the high redshift observations at face value, then, our
results confirm that massive galaxies, both individually and as a
population, must undergo considerable structural evolution over the
interval $z \lesssim 2.3$ in order to develop into the kinds of
galaxies seen locally---even after star formation in these galaxies
has effectively ended.  We see some hints that a significant amount of
this evolution ($\lesssim 50$ \%) may have already occurred by $z \sim
1.6$.

\vspace{0.2cm}

The fact that each and every one of the \citet{VanDokkum} galaxies
must undergo significant structural evolution to match the properties
of present-day galaxies implies that the mechanism that drives this
growth must apply more or less evenly to all galaxies.  To see this,
let us assume that some external process drives the size evolution of
these galaxies, and that even a single event is sufficient to move an
individual galaxy onto the main size--mass relation.  Then, we can
assume some simple probability distribution for the number of events,
$N$, among individual galaxies.  (For example, we could assume that
events occur randomly across the time interval $z < 2.3$, or that each
galaxy experiences $N \pm \sqrt{N}$ events.)  Now, our results suggest
that the number density of \citet{VanDokkum}--like galaxies drops by
at least a factor of 5000 since $z \sim 2.3$.  In order to ensure that
at most 1/5000 galaxies have $N = 0$ after $z \sim 2.3$, simple
probabilistic arguments imply that the average galaxy must undergo
$\gtrsim 20$ events.  This would imply that a strongly stochastic
process like major mergers cannot be the primary mechanism for the
strong size evolution of massive galaxies.

Apart from their small sizes and high velocity dispersions, our
compact galaxy candidates are not obviously distinct from the general
population (Figure \ref{fig:props}).  If anything, at fixed velocity
dispersion, our compact galaxies have stellar populations that are
slightly younger than average (at $\sim 1.9 \sigma$ significance).
Even so, the majority of these galaxies' stellar populations are
definitely `old', with luminosity-weighted mean stellar ages typically
in the range 6--10 Gyr.  But if some external mechanism drives the
size evolution of these galaxies, we speculate that their small sizes
may indicate that they have assumed their present form comparatively
{\em early}, and in this sense they may actually be relatively {\em
old} \citep[see also, \eg ,][]{VanDerWel2009}.  If so, with better
understanding of the processes that determine the sizes of early type
galaxies, and in particular the role of merging, the properties of
these galaxies could provide a means of constraining the evolution of
massive galaxies after they have completed their star formation,
including their late-time merger histories.

\begin{appendix}

\section{Looking for Massive Compact Galaxies in the SDSS Photometric Sample}
\label{ch:photsample}

In this Appendix, we present a complementary analysis in which we
directly compare the spectroscopic and photometric samples, in order
to test the conclusion that the lack of massive, compact galaxies in
the spectroscopic sample cannot be explained by the 
selection effects.

\begin{figure*}
\includegraphics[width=17.5cm]{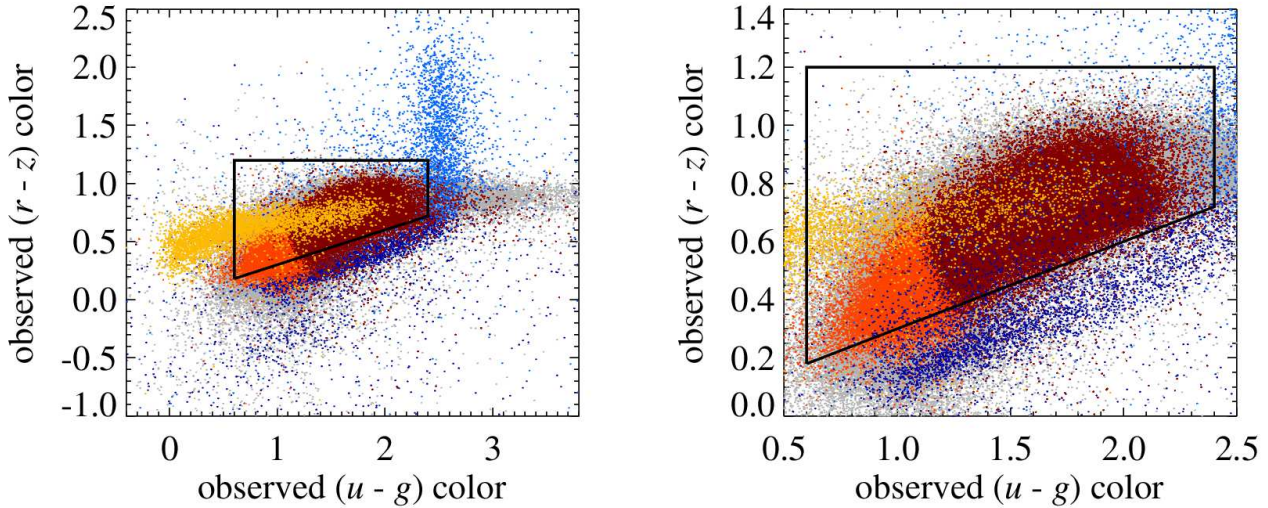}
	\caption{Selecting $z \lesssim 0.1$ galaxies based on color
        alone. Each panel shows the observed ($u-g$)--($r-z$)
        color--color plot for objects in the SDSS spectroscopic
        sample; the right panel simply shows the central region in
        greater detail. Points are color--coded according to their
        spectral classification, \viz : galaxies ({\em grey}),
        galaxies with $0.066 < z < 0.10$ ({\em light red}), galaxies
        with $0.066 < z < 0.10$ and $^{0.1}(u-r) > 2.5$ ({\em dark
        red}), quasars ({\em yellow}), late-type stars ({\em light
        blue}), and ordinary stars ({\em dark blue}), . The box shows
        the color selection that we use to select $z \lesssim 0.1$
        galaxies. This selection should produce a reasonably complete
        sample of $z \lesssim 0.1$ galaxies, with some contamination
        from both stars and quasars.  In particular, later-type stars
        and some quasars have observed SEDs that are very similar to
        red sequence galaxies at $z \sim 0.1$. \label{fig:colselect}}
\end{figure*}

\subsection{Selecting Galaxies by Color Alone} \label{ch:colsel}

Before we can address the question of massive compact galaxies in the
SDSS photometric sample, we must first devise a means of separating
stars and galaxies without selecting on the basis of observed size or
light profile. Our method for doing so is shown in
Figure~\ref{fig:colselect}, which plots the observed
(extinction-corrected) $ugrz$ colors of different classes of objects
from the spectroscopic sample; we show: all galaxies (grey), $0.066 <
z < 0.12$ galaxies (bright red), and those with $^{0.1}(u-r) > 2.5$
(dark red), O--K stars (dark blue), M-type or later stars (light
blue), and quasars (yellow).

The black box shown in Figure~\ref{fig:colselect} shows our criteria
for selecting $0.066 < z < 0.12$ galaxies based on their $ugrz$ colors:
\begin{eqnarray}  0.6 & < &(u-g) < 2.4 ~~~~~ \mathrm{and}  \nonumber \\ 
	0.3 \times (u-g) & < &(r-z) < 1.2 ~ . 
\end{eqnarray} 
Again, we apply this selection in terms of model colors. Note how,
whereas the stellar sequence is reasonably well separated from the
region of color space occupied by galaxies for $(u-g) \lesssim 2.5$,
beyond this point, the late-type stellar sequence turns up, such that
late-type stars and galaxies are blended. In the most general terms
possible, the mean galaxy redshift increases towards redder $(u-g)$
colors. This means that our ability to distinguish red galaxies from
late-type stars on the basis of their optical SEDs is limited to $z
\lesssim 0.12$.

In the right-hand panel of Figure~\ref{fig:colselect}, we zoom in on
this selection region. From this panel, it is clear that a large
proportion of quasars will also be included in our color-selected
`galaxy' sample. Similarly, it is clear that this color selection is
not 100 \% efficient in excluding stars from our sample: more
quantitatively, with this selection we are able to exclude more than 80
\% of spectroscopically identified stars that are given $0.066 < \zphot
< 0.12$, while retaining more than 97 \% of all $0.066 < \zphot < 0.12$
galaxies. Furthermore, it should be remembered that stars are already
heavily selected against for the spectroscopic sample plotted in
Figure~\ref{fig:colselect}; the relative number of stellar
`contaminants' may well be considerably higher for the photometric
sample.

\subsection{Photometric Redshifts and Stellar Mass Estimates}

A major improvement in DR7 is a complete revision in how the basic
(\texttt{photoz}) photometric redshifts are derived
\citep{Abazajian2009}. Rather than using some combination of synthetic
template spectra to reproduce the observed colors of individual
galaxies, the new \texttt{photoz} algorithm directly compares the
observed photometry of individual galaxies to that of galaxies that
have spectroscopic redshifts. Specifically, for each individual
object, the algorithm finds the 100 closest neighbours in $ugriz$
color space, and fits a hyper-plane to these points, rejecting
outliers; the redshift is then determined by interpolating along this
4D surface. In comparison to the DR6 algorithm, this reduces the RMS
redshift error by more than 75 \% ($\left<\Delta z\right>$ = 0.025),
and significantly reduces systematic errors \citep{Abazajian2009}.

For this analysis, rather than full SED-fit stellar mass estimates
assuming the photometric redshifts, we will simply make use of the
empirical relation between $^{0.1}(g-i)$ color and $M_*/L$ (Equation
\ref{eq:mls}).  In this way, we are able to recover the
$\zspec$-derived, SED-fit $M_*/L$s of the sample of galaxies shown in
Figure \ref{fig:mls} to 0.045 dex ($1\sigma$); including the effects
of photometric redshift errors, k-corrections, and $M_*/L$ errors, the
total ($1\sigma$) error in $M_*$ is 0.13 dex. This should be compared
to the median formal error on the original SED-fit stellar mass
estimates, 0.10 dex; that is, the errors on $M_*$ based on photometric
masses (adding these two errors in quadrature) are only about 60 \%
greater than those based on spectroscopic redshifts.

\subsection{The Size Distribution of Massive, Red Sequence Galaxies} \label{ch:sizemass3}

\begin{figure}
\includegraphics[width=7.9cm]{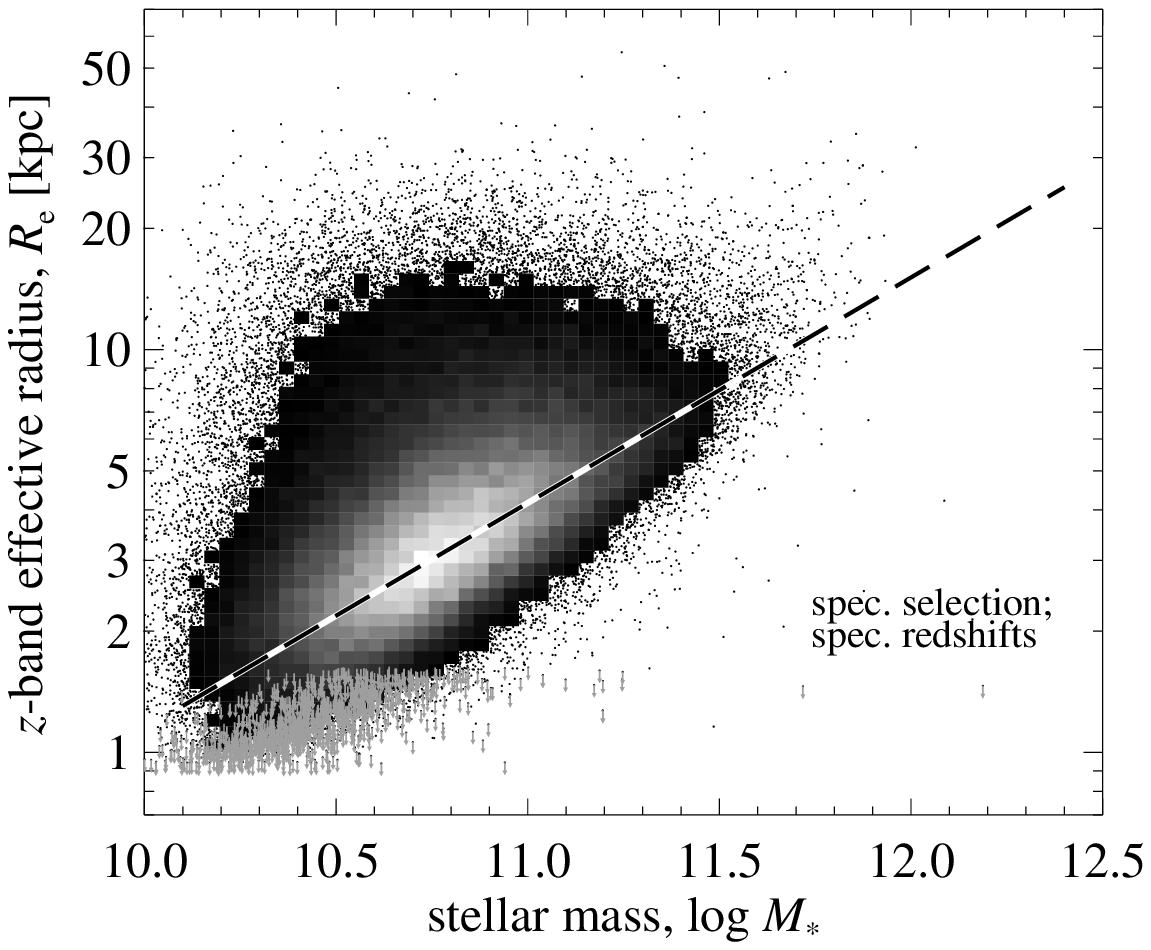}
\includegraphics[width=7.9cm]{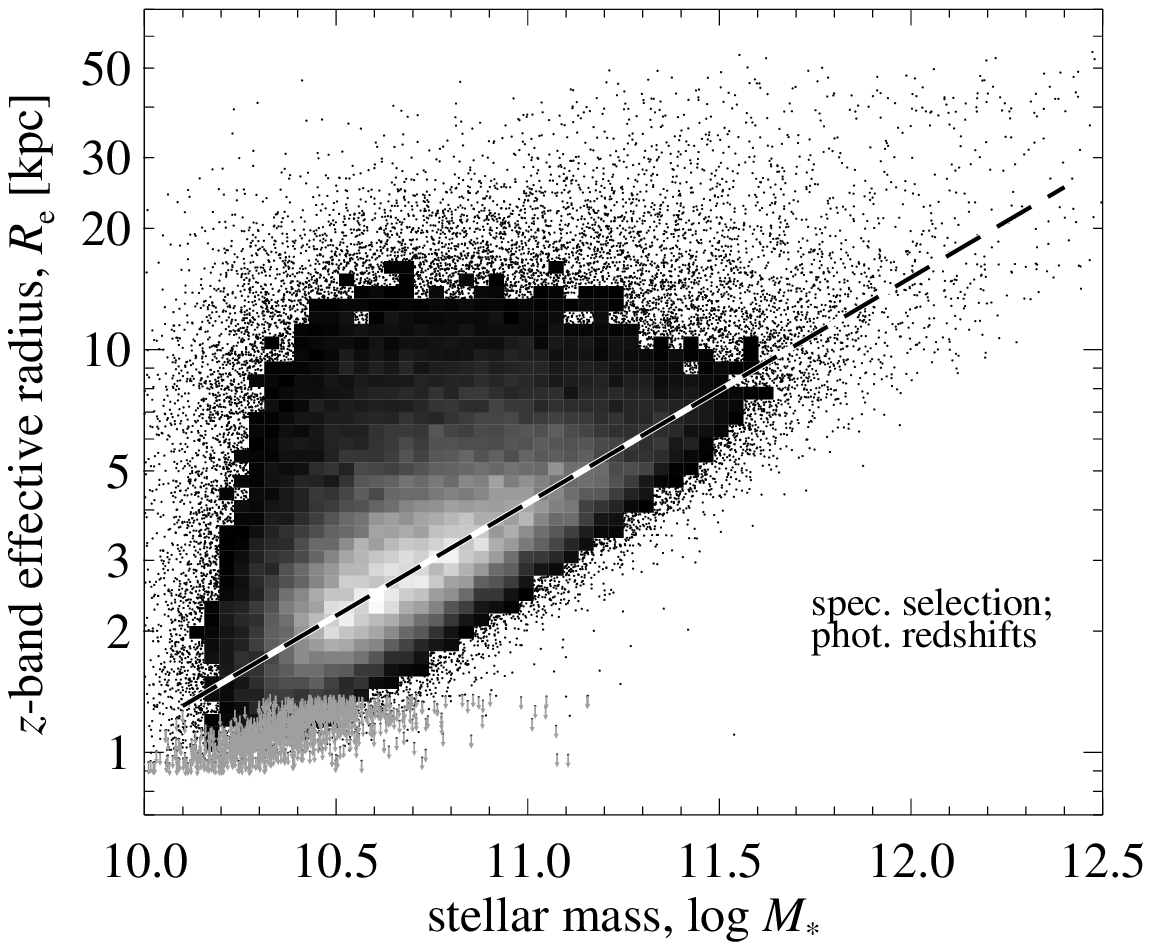}
\includegraphics[width=7.9cm]{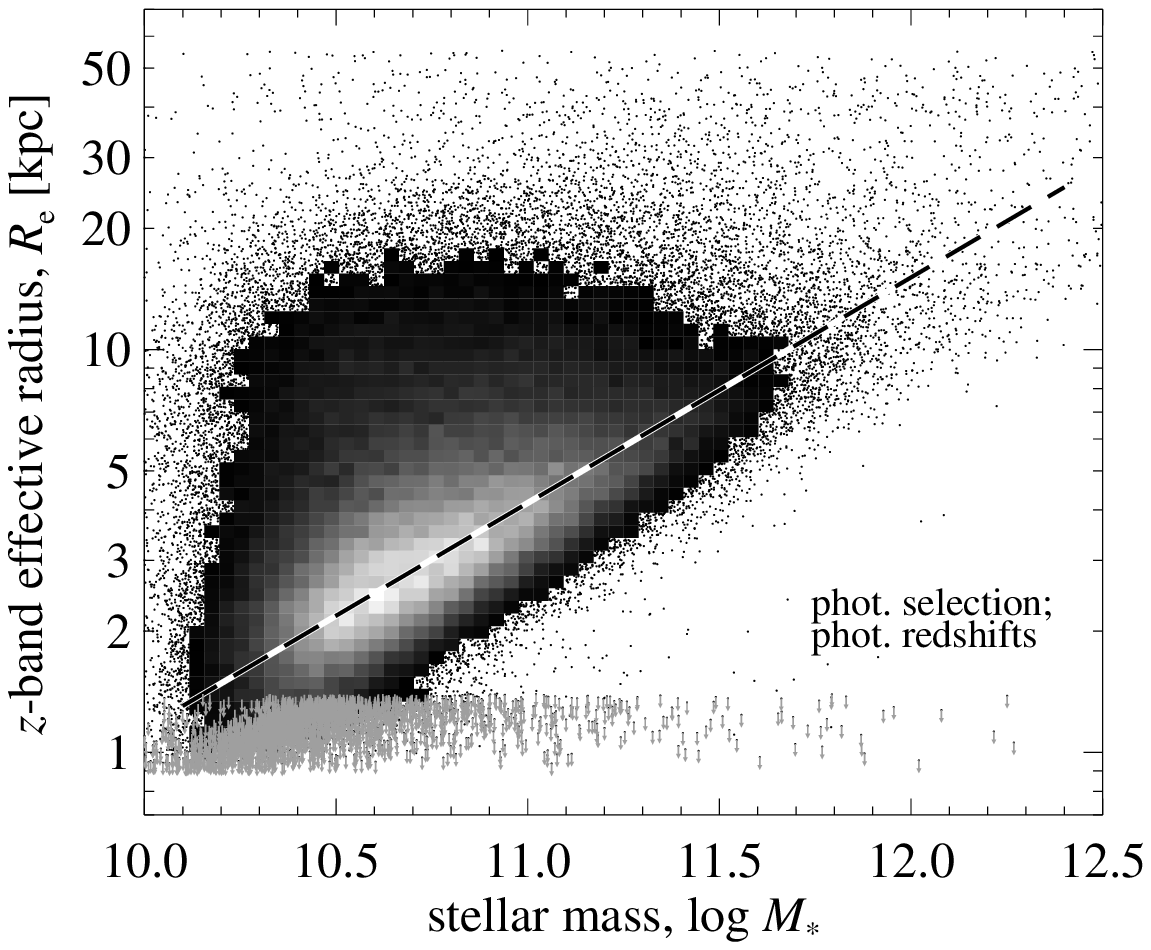}
\caption{The size--mass plot for massive, red sequence galaxies at
  $0.066 < z < 0.12$, based on the spectroscopic and the full
  photometric SDSS catalogs. Each panel shows the sizes and masses
  of galaxies based on, from top to bottom, the spectroscopic sample
  using spectroscopic redshifts, the spectroscopic sample using
  photometric redshifts, and the photometric sample using photometric
  redshifts; in each case, only those objects inferred to have
  $^{0.1}(u-r) > 2.5$, and $0.066 < z < 0.10$ are shown.  In panel 3,
  many more objects with inferred sizes $\lesssim 0.3$ kpc can be
  seen; these are largely stars misclassified (in terms of their
  photometric redshifts) as galaxies. For $M_* \lesssim 10^{10.8}$
  M\sol\ and $R_{\mathrm{e}} \lesssim 10^{-0.2}$ kpc, comparison
  between panels 2 and 3 suggest that there may be a few additional
  galaxies in the photometric sample that do not appear in the
  spectroscopic sample.
        \label{fig:sizemass3}}
\end{figure}

\begin{figure} \centering
\includegraphics[width=8.6cm]{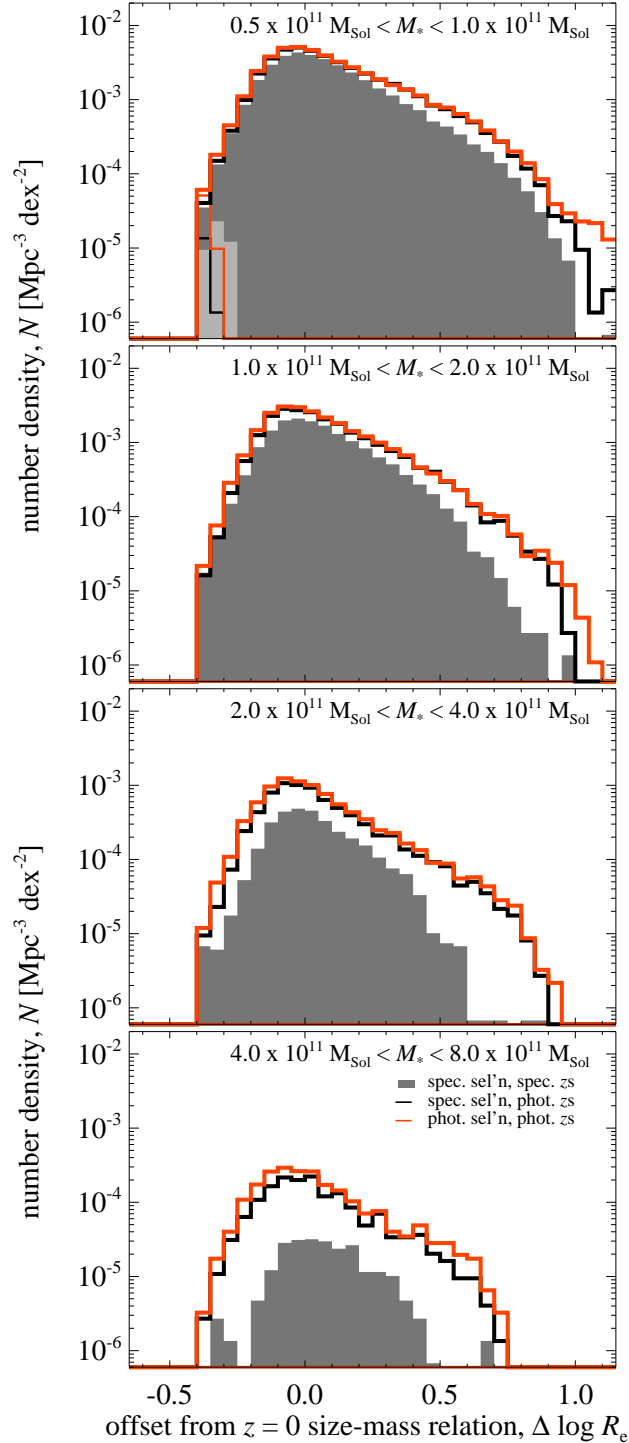}
	\caption{The observed size distribution of massive, red
        galaxies at $0.066 < z < 0.12$; each panel is for a different
        mass range as marked. In each panel, the solid histogram
        represents the SDSS spectroscopic sample, analyzed using
        spectroscopic redshifts.  The black and red histograms are the
        SDSS spectroscopic and photometric samples, respectively,
        analyzed using photometric redshifts.  We have visually
        inspected all objects with inferred $M_* > 10^{11}$ M\sol\ and
        $\Delta \reffz < -0.5$ dex; not one of these objects is a
        plausible massive, comapct galaxy candidate.  The fact that
        the shape of the red histogram does not differ significantly
        from that of the black histogram for $\Delta \reffz > -0.5$
        dex indicates that the spectroscopic sample is not
        significantly biased against compact galaxies.
	\label{fig:sizedists2}}
\end{figure} 

In Figure \ref{fig:sizemass3}, we show three size--mass diagrams
corresponding to, from top to bottom: ({\em a}.)\ the spectroscopic
sample, analyzed using spectroscopic redshifts; ({\em b}.)\ the
spectroscopic sample, analyzed using photometric redshifts; and ({\em
c}.)\ the photometric sample, analyzed using photometric redshifts.
In all three cases, the only selections applied to each sample are on
photometric \texttt{type} (to exclude optical artifacts, etc., we
require either a star or galaxy \texttt{type} classification) and
$ugrz$ color (to exclude stars); then, as in
Figures~\ref{fig:sizemass} and \ref{fig:sizecomp}, we are only showing
those galaxies inferred to have $0.066 < z < 0.12$ and $^{0.1}(u-r) >
2.5$.  Again, objects with measured sizes smaller than $0\farcs75$ are
shown as upper limits, assuming a size of $0\farcs75$. When comparing
these three different analyses, the difference between ({\em a}.)\ and
({\em b}.)\ shows the effect of using spectroscopic versus photometric
redshifts, and the difference between ({\em b}.)\ and ({\em c}.)\
shows the difference between the SDSS spectroscopic and photometric
selection.  That is, the comparison between ({\em b}.)\ and ({\em
c}.)\ gives a direct indication of the level of incompleteness in the
spectroscopic sample.

Looking at panels ({\em a}.)\ and ({\em b}.), it is clear that the use
of photometric redshifts produces a considerably greater scatter in
the size--mass diagram, including a rather large number of galaxies
with inferred stellar masses of $10^{12}$ M\sol\ or greater.  There is
a clear excess of unresolved objects with inferred stellar masses
greater than $\sim 10^{11}$ M\sol\ in panel ({\em b}.)\ in comparison
to panel ({\em a}.) However, we already know from section
\ref{ch:specsample} that there are no objects in the spectroscopic
sample with these sizes and masses---these objects cannot be genuine
compact galaxies. Of the 34 with inferred $M_* > 10^{11}$ M\sol , 16
of these objects are spectrally identified as being stars, and one as
a quasar at $z=0.102$.  Of the 17 spectrally confirmed galaxies, all
have $|z_\mathrm{phot} -z_\mathrm{spec}| \gtrsim 0.02$.  Of these, 15
have had their redshifts, and thus stellar masses, seriously
overestimated; the other two are at $z > 0.12$, and so have had their
intrinsic sizes underestimated.

Turning now to the comparison between panels ({\em b}.)\ and ({\em
c}.), the first point to make is that the excess of unresolved sources
is even more pronounced.  We have matched all of these objects to the
2MASS point source catalog in order to investigate their NIR colors.
90 \% of these objects fall in the stellar region of the $(J-K)$--$K$
color--magnitude plot; similarly, 80 \% fall in the stellar region of
a $(g-z)$--$(J-K)$ color--color plot.

Further, we have visually inspected the 434 objects with inferred $M_*
> 10^{11}$ M\sol\ and with sizes smaller than the main $M_*$--$\reffz$
relation by $0.4$ dex or more.  Roughly 70 \% of these objects are
obviously stars: 133 come from crowded Galactic fields covered as part
of SEGUE; 126 are double stars; 49 have clear diffraction spikes
and/or are clearly saturated.  Another 12 objects have been
cross-matched with the USNO-B star catalog (within $1''$), and have
measured proper motions of 1---4$''$/yr.  19 objects are the central
point sources of very large spiral galaxies; most of these are also
found in the ROSAT and/or FIRST catalogs.  We also note that there are
17 very small disk or irregular galaxies with red point sources at or
very near their centers. Most of these also have proper motion
measurements from the USNO-B catalog, and several are spectrally
identified as late type stars; it seems plausible that these galaxies
simply have foreground stars coincidentally superposed very near their
centers.

In short, of the 434 objects from the full photometric sample that, on
the basis of photometric redshifts, are inferred to have $M_* >
10^{11}$ M\sol\ and $\Delta \reffz < -0.4$ dex, not one remains as a
viable compact galaxy candidate.

\subsection{Estimating the Importance of Spectroscopic Selection Effects}

The conclusion from both the analyses that we have now presented is
that there are no galaxies in the local universe with sizes and masses
comparable to the compact galaxies found at higher redshifts.  In
Figure \ref{fig:sizedists2}, we provide a more quantitative statement
of this conclusion, by plotting the size distribution for massive, red
galaxies in different mass bins.

In this figure, the filled histograms represent the main SDSS
spectroscopic sample, analyzed using spectroscopic redshifts, as in
\textsection\ref{ch:specsample}.  The heavy black and red histograms
represent the spectroscopic and photometric samples, respectively,
analyzed using photometric redshifts, as in
\textsection\ref{ch:sizemass3}.  In all cases, objects excluded on the
basis of visual inspection are not included; this accounts for the
sharp cutoffs at $\Delta\log\reffz = -0.3$ and at $\Delta\log\reffz =
-0.4$ for the filled and open histograms, respectively.  Immediately
above these cutoffs, where we have not visually inspected individual
objects, but where there is likely to still be significant
contamination, these distributions should be regarded as upper limits
on the true distribution.  In the upper panel, we plot those objects
with observed sizes smaller than $0\farcs75$ separately as the light
grey filled histogram, and the thin black and red histograms.

As in \textsection\ref{ch:sizemass3}, the difference between the
filled and solid black histogram, both of which are derived from the
spectroscopic sample, shows the increased scatter due to the use of
photometric redshifts.  

Similarly, the difference between the black and red histograms show
the difference between the spectroscopic and photometric samples, and
so allow a quantification of the bias in the spectroscopic sample.  By
simply tallying the numbers of galaxies with $-0.4 < \Delta\log\reffz
< -0.3$, we find that the `completeness' (the ratio between the number
of galaxies in the spectroscopic sample compared to the full
photometric sample) is 75 \%, 68 \%, 67 \%, and 43 \% for each of
these mass bins, from lowest to highest.

In order to improve on these estimates, we have done the following.
Using the approach described above, we have assigned each object a
weight according to its $\zphot$--derived mass and size.  Then, going
back to the spectroscopic sample, we use these to compute the mean
weight in cells of $\zspec$--derived mass and size.  The completeness
contours we derive in this way are in good qualitative agreement with
those shown in Figure \ref{fig:sizemass}, although they suggest
incompleteness at the 2--5 \% level for mean--sized galaxies with $M_*
\gtrsim 10^{11}$ M\sol .  Using these values to estimate the number of
$M_* > 10^{11}$ M\sol\ galaxies with $\Delta\reffz < -0.3$ dex, this
suggests that the spectroscopic sample is missing on the order of 4
such galaxies.

\end{appendix}

\end{document}